\documentclass[11pt]{article}
\pdfoutput=1

\usepackage{jheppub}
\usepackage{amssymb,amsfonts,amsmath,amsthm, enumerate}
\usepackage{mathrsfs}
\usepackage{tikz}
\usetikzlibrary{shapes.geometric, arrows}
\usetikzlibrary{arrows.meta}
\usepackage{comment}
\usepackage{hyperref}

\usepackage{datetime2}

\usepackage{arydshln}

\usepackage{calc}

\usepackage{accents}

\makeatletter
\newcommand{\Vast}{\bBigg@{4.75}}
\makeatother

\usepackage{cancel}

\newcommand{\CD}{\mathcal{D}}

\newcommand{\hA}{{\hat{A}}}
\newcommand{\hB}{{\hat{B}}}
\newcommand{\hmu}{i}
\newcommand{\hnu}{j}

\newcommand\qt\tau

\definecolor{darkgreen}{rgb}{0.0, 0.4, 0.1}

\newcommand{\p}{\partial}

\renewcommand{\tilde}[1]{\widetilde{#1}}

\makeatletter
\renewcommand{\@seccntformat}[1]{\csname the#1\endcsname.\,\,}
\makeatother
\allowdisplaybreaks

\let \savenumberline \numberline
\def \numberline#1{\savenumberline{#1.}}

\makeatletter
\def\@fpheader{\relax}
\makeatother

\def\bea{\begin{eqnarray}}
\def\eea{\end{eqnarray}}

\usetikzlibrary{shapes,arrows,chains}
\usetikzlibrary{decorations.markings}
\usetikzlibrary{decorations.pathmorphing}
\usetikzlibrary{shapes.multipart}
\tikzset{snake it/.style={decorate, decoration=snake}}

\newcommand{\RR}{\mathbb{R}}

\newcommand{\pd}{\partial}

\newcommand{\mD}{\mathcal{D}}

\newcommand{\OO}{\mathcal{O}}

\newcommand{\mJ}{\mathcal{J}}
\newcommand{\mC}{\mathcal{C}}
\newcommand{\mH}{\mathcal{H}}

\def\mN{\mathcal{N}}

\def\mD{\mathcal{D}}
\def\mC{\mathcal{C}}

\def\mH{\mathcal{H}}

\usepackage{graphbox}

\title{\ \vspace{1.6cm} \\
Aspects of Nonrelativistic Strings}
\author[a,b]{Gerben Oling}
\author[b]{and Ziqi Yan}
\affiliation[a]{%
  The Niels Bohr Institute,
  University of Copenhagen,
  \protect\\
  Blegdamsvej 17,
  DK-2100 Copenhagen Ø,
  Denmark\medskip
}
\emailAdd{gerben.oling@su.se}
\emailAdd{ziqi.yan@su.se}
\affiliation[b]{%
  Nordita,
  KTH Royal Institute of Technology and Stockholm University,
  \protect\\
  Hannes Alfv\'{e}ns v\"{a}g 12,
  SE-106 91 Stockholm,
  Sweden
}
\abstract{%
We review recent developments on nonrelativistic string theory.
In flat spacetime, the theory is defined by a two-dimensional relativistic quantum field theory with nonrelativistic global symmetries acting on the worldsheet fields.
This theory arises as a self-contained corner of relativistic string theory.
It has a string spectrum with a Galilean dispersion relation, and a spacetime S-matrix with nonrelativistic symmetry.
This string theory also gives a unitary and ultraviolet complete framework that connects different corners of string theory, including matrix string theory and noncommutative open strings.
In recent years, there has been a resurgence of interest
in the non-Lorentzian geometries and quantum field theories that arise from nonrelativistic string theory in background fields.
In this review, we start with an introduction to the foundations of nonrelativistic string theory in flat spacetime.
We then give an overview of recent progress, including the appropriate target-space geometry that nonrelativistic strings couple to.
This is known as (torsional) string Newton--Cartan geometry, which is neither Lorentzian nor Riemannian.
We also give a review of nonrelativistic open strings and effective field theories living on D-branes.
Finally, we discuss applications of nonrelativistic strings to decoupling limits in the context of the AdS/CFT correspondence.
}

\dedicated{Review article for Frontiers in Physics Research Topic \href{https://www.frontiersin.org/research-topics/19214/non-lorentzian-geometry-and-its-applications}{Non-Lorentzian Geometry and its Applications}}

\begin{document}

\maketitle

\section{Introduction}
\label{sec:intro}
It has long been known that different string theories are limits of M-theory.
While the various corners in this web that are described by perturbative string theories are fairly well understood, we are still far from a complete understanding of nonperturbative regimes in the full M-theory.
For example, exploring nonperturbative aspects of string/M-theory is important for understanding the information paradox for black holes, which are fundamentally nonperturbative objects.
One nonperturbative approach to M-theory stems from taking a subtle limit of the compactification on a spacelike circle.
This notably leads to Matrix theory~\cite{Banks:1996vh, Susskind:1997cw, Seiberg:1997ad, Sen:1997we, Hellerman:1997yu, deWit:1988wri},
which serves as a powerful tool for understanding the full M-theory in a simple system of D0-branes.

Similarly, by taking an infinite boost limit of the compactification of string theory on a spacelike circle, we are led to the discrete light cone quantization (DLCQ) of strings,
which has a Matrix string theory description \cite{Motl:1997th, Banks:1996my, Dijkgraaf:1997vv}.
The infinite boost limit along a spacelike circle can be interpreted as a compactification on a lightlike circle, which leads to nonrelativistic~(NR) behavior in the resulting frame (see for example \cite{Bigatti:1997jy}).
From a different perspective, it is known that the DLCQ of string theory arises from a T-duality transformation along a compactified spacelike circle in a genuine NR theory~\cite{Klebanov:2000pp, Gomis:2000bd, Danielsson:2000gi}.
This theory is a unitary and ultraviolet (UV) complete string theory described by a two-dimensional quantum field theory (QFT) with a Galilean-like global symmetry in flat spacetime. This NR symmetry is realized by introducing extra one-form worldsheet fields in addition to the ones that are target-space coordinates.
The theory has a spectrum of string excitations that satisfy a `string' Galilean-invariant dispersion relation, and hence it has a spacetime S-matrix with NR symmetries.
For these reasons, such a theory is referred to as \emph{nonrelativistic string theory} in the literature \cite{Gomis:2000bd}.~\footnote{Also see \cite{Danielsson:2000gi}, where NR string theory is referred to as `wound string theory.'
In \cite{Gomis:2000bd}, a no-ghost theorem similar to the one in relativistic string theory has been put forward for NR string theory, showing unitarity of the theory.
Moreover, tree-level and one-loop NR closed string amplitudes have been studied in \cite{Gomis:2000bd, Danielsson:2000gi}, showing that NR string theory is a self-consistent, UV-finite perturbation theory in the genus. Higher-genus amplitudes have also been discussed in \cite{Gomis:2000bd}. We will not focus on string amplitudes in this review, but a brief discussion can be found at the end of \S\ref{ssec:flat-nr-closed-strings}.}
Via T-duality, NR string theory provides a microscopic definition of string theory in the DLCQ, which is otherwise only defined as a subtle
limit.
In the formalism of NR string theory, the exotic
physics of string/M-theory in the DLCQ with compactification on a lightlike circle is now translated to the more familiar language of NR physics.

There are no massless physical states in NR string theory, and the associated low-energy effective theory is described by a Newton-like theory of gravity, instead of General Relativity~\cite{Gomis:2000bd, Danielsson:2000gi}.
Since it is UV finite, NR string theory provides a UV completion of the associated theory of gravity in the same way that relativistic string theory provides a UV completion of Einstein's gravity \cite{Gomis:2000bd, Danielsson:2000gi, Danielsson:2000mu}.
In this sense, NR string theory defines a NR theory of quantum gravity.
As such, it provides us with a novel approach towards understanding relativistic quantum gravity, orthogonal and hopefully complementary to the usual paths towards quantum gravity that start from relativistic classical gravity or relativistic QFT.

Recently, there has been renewed interest in NR string theory, based for a large part on understanding the precise notion of its target space geometry, starting with the early work in~\cite{Andringa:2012uz}.
The appropriate geometry that NR strings couple to is now known as \emph{string Newton--Cartan} geometry, which is neither Riemannian nor Lorentzian.
The original notion of Newton--Cartan geometry was introduced to geometrize Newtonian gravity, and hence only distinguishes a single direction which is associated to time.
In contrast, string Newton--Cartan geometry generalizes this notion to distinguish two directions that are longitudinal to the string.

We start this review in \S\ref{sec:wnrst} by introducing the defining action of NR string theory in flat spacetime.
We review how this string theory is embedded in relativistic string theory as a decoupling limit, where parts of the spectrum decouple and the remaining states satisfy a Galilean-invariant dispersion relation.
This is achieved by coupling winding relativistic string states to a background Kalb-Ramond field, which is fine tuned such that its energy cancels the string tension.
We elaborate on basic ingredients of NR closed and open strings, and review how they are related to relativistic strings in the DLCQ via T-duality.
In \S\ref{sec:nscs}, we review recent progress on classical NR strings in curved spacetime.
This leads to (torsional) string Newton-Cartan geometry in the target space.
In \S\ref{sec:eftns}, we discuss quantum aspects of the sigma model for NR strings in curved spacetime.
We will also review different target-space effective theories that arise from imposing the worldsheet Weyl invariance at quantum level.
Next, in \S\ref{sec:nhd}, we discuss applications of NR strings to the AdS/CFT correspondence.
We focus on a limit of NR string theory that results in sigma models with a NR worldsheet.
These theories are related to decoupling limits of AdS/CFT that lead to Spin Matrix theories.
In~\S\ref{sec:co}, we conclude the review and comment on other interesting lines of research in the field.

Finally, it is important to point out that several different limits of string theory that lead to NR symmetries have been considered in the literature.
We will always use the term `nonrelativistic string theory' to refer to the theory we mentioned above, but some of the other approaches are sketched in \S\ref{sec:co}.

\section{What is Nonrelativistic String Theory?} \label{sec:wnrst}
We start with reviewing the sigma model describing nonrelativistic (NR) string theory in flat spacetime with a NR global symmetry that was first introduced in~\cite{Gomis:2000bd}.
We will review its basic ingredients, how it arises from relativistic string theory, and its relation to other corners of string theory.

\subsection{Nonrelativistic string theory in flat spacetime}
\begin{figure}[t]
\begin{minipage}{6in}
  \centering
  \includegraphics[align=c,height=2in]{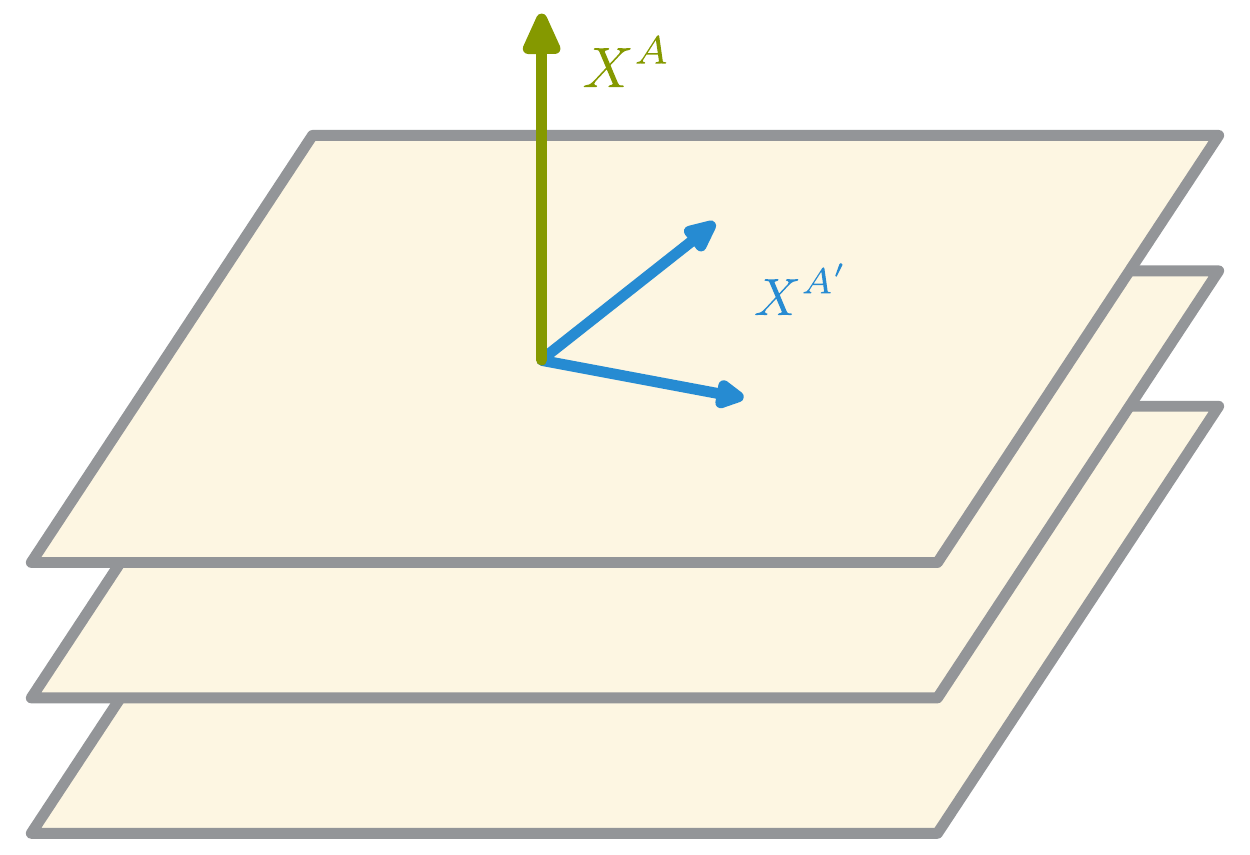}
  \hspace*{.2in}
  \includegraphics[align=c,height=2in]{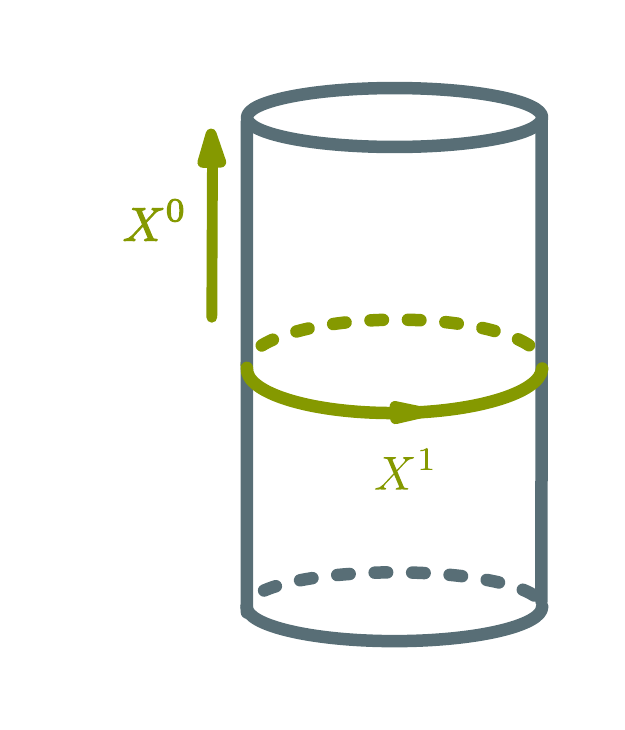}
\end{minipage}
    \caption{A schematic illustration of the longitudinal and transverse fields $X^A$ and $X^{A'}$ in the NR string action.
    The green arrows represent the two longitudinal directions $X^A$.
    Each horizontal slice represents the $(d-2)$-dimensional transverse directions $X^{A'}$.}
    \label{fig:flat-bread}
\end{figure}

In this review, we work with a Euclidean worldsheet corresponding to a Riemann surface~$\Sigma$, parametrized by $\sigma^\alpha = (\sigma^1,\sigma^2)$, where $\sigma^2$ is the Euclidean time.
We denote the worldsheet metric and the worldsheet zweibein by $h_{\alpha\beta}$ and $e_\alpha{}^a$\,, with $a = 1, 2$\,, such that $h_{\alpha\beta} = \delta_{ab} \, e_\alpha{}^a \, e_\beta{}^b$.
The worldsheet fields consist of the scalars $X^\mu = (X^0, \cdots, X^{d-1})$ that map $\Sigma$ to a $d$-dimensional spacetime manifold $\mathcal{M}$\,, and in addition two one-forms that we denote by $\lambda$ and~$\bar{\lambda}$\,.
The worldsheet scalars $X^\mu$ play the role of spacetime coordinates.
In NR string theory, two \emph{longitudinal} spacetime coordinates are distinguished from the remaining $d-2$ \emph{transverse} coordinates, as illustrated in Figure~\ref{fig:flat-bread}.
These directions are denoted by $X^A=(X^0,X^1)$ and $X^{A'}=(X^2,\ldots,X^{d-1})$, respectively.
The defining action for NR string theory in flat spacetime is \cite{Gomis:2005pg, Bergshoeff:2018yvt}
\begin{equation} \label{eq:Sfree}
    S = \frac{1}{4\pi\alpha'} \int_\Sigma d^2 \sigma \sqrt{h} \, \Bigl( h^{\alpha\beta} \, \p_\alpha X^{A'} \, \p_\beta X_{A'} + \lambda \, \bar{\mathcal{D}} X + \bar{\lambda} \, \mathcal{D} \overline{X} \Bigr)\,,
\end{equation}
where $\alpha'$ is the Regge slope and $h = \det h_{\alpha\beta}$\,.
Transverse indices are lowered using the flat Euclidean metric $\delta_{A'B'}$, whereas the longitudinal directions contain a Minkowski structure.
We introduced the light-cone coordinates $X$ and $\overline{X}$ in the target-space longitudinal sector and the worldsheet derivatives $\CD$ and $\bar{\CD}$\,,
\begin{subequations}
\begin{align}
    X & = X^0 + X^1,
        &%
    \mathcal{D} & = i \, h^{-1/2} \, \epsilon^{\alpha\beta} \, (e_\alpha{}^1 + i \, e_\alpha{}^2) \, \p_\beta\,,
        \\[2pt]
    \overline{X} & = X^0 - X^1,
        &%
    \bar{\mathcal{D}} & = i \, h^{-1/2} \, \epsilon^{\alpha\beta} \, (- e_\alpha{}^1 + i \, e_\alpha{}^2) \, \p_\beta\,.
\end{align}
\end{subequations}
Here, the worldsheet Levi-Civita symbol $\epsilon^{\alpha\beta}$ is defined by $\epsilon^{12} = +1$\,.
In conformal gauge, we set $h_{\alpha\beta} = \delta_{\alpha\beta}$ so that $\mathcal{D}=\pd=\pd_1+i\pd_2$ and $\bar{\mathcal{D}}=\bar\pd=\pd_1-i\pd_2$\,, and the action~\eqref{eq:Sfree} becomes
\begin{equation} \label{eq:cg}
    S = \frac{1}{4\pi\alpha'} \int_\Sigma d^2 \sigma \, \Bigl( \p_\alpha X^{A'} \, \p^\alpha X_{A'} + \lambda \, \bar{\p} X + \bar{\lambda} \, \p \overline{X} \Bigr)\,,
\end{equation}
which is also known as the \emph{Gomis--Ooguri} string theory~\cite{Gomis:2000bd}.

In conformal gauge, the fields $\lambda$ and $\bar{\lambda}$ transform~\cite{Gomis:2000bd, Gomis:2019zyu} under the worldsheet diffeomorphism parametrized by $\xi^\alpha$
as $\delta \lambda = \xi \, \p \lambda + \lambda \, \p \xi$ and $\delta \bar{\lambda} = \bar{\xi} \, \bar{\p} \bar{\lambda} + \bar{\lambda} \, \bar{\p} \bar{\xi}$\,,
where $\xi = \xi^0 + \xi^1$ and $\bar{\xi} = \xi^0 - \xi^1$\,.
This implies that $\lambda$ and $\bar{\lambda}$ transform as (1,0)- and (0,1)-forms, respectively.
In the action~\eqref{eq:cg}, they are Lagrange multipliers that impose the chirality conditions
\begin{equation}
  \label{eq:holocond}
  \bar{\p} X = \p \overline{X} = 0\,,
\end{equation}
on the longitudinal directions.\, \footnote{The quantum mechanical implementation of the constraints \eqref{eq:holocond} in string loops will be reviewed in \S\ref{ssec:flat-nr-closed-strings}.}
The global symmetry algebra of the NR string action~\eqref{eq:cg} consists of an infinite number of spacetime isometries~\cite{Batlle:2016iel}.
This algebra contains two copies of the Witt algebra, which are related to the (anti-)holomorphic reparametrizations associated to the constraints~\eqref{eq:holocond}.
It also includes a Galilei-like boost symmetry
that acts on the worldsheet fields $X^\mu$ as,
\begin{equation}
    \delta_{\text{G}} X^A = 0\,,
        \qquad
    \delta_{\text{G}} X^{A'} = \Lambda^{A'}{}_A X^A\,,
\end{equation}
which is referred to as the \emph{string Galilei boost} symmetry.
This is a natural generalization of the Galilei boost symmetry for NR particles: while the Galilei boost acts differently on space and time directions, string Galilei boosts act differently on the directions longitudinal and transverse to the string.
Additionally, for the action \eqref{eq:cg} to be invariant under string Galilei boosts, the one-form fields are required to transform as follows:
\begin{equation}
    \delta_{\text{G}} \lambda = \bigl( \Lambda_0{}^{A'} + \Lambda_1{}^{A'} \bigr) \, \p X_{A'},
        \qquad
    \delta_{\text{G}} \bar{\lambda} = \bigl( \Lambda_0{}^{A'} - \Lambda_1{}^{A'} \bigr) \, \bar{\p} X_{A'}.
\end{equation}
This implies that the two-dimensional QFT defined by the action~\eqref{eq:cg} has a NR global symmetry that acts on worldsheet fields.
Consequently, as we will see in \S\ref{sec:zeroRegge}, this theory
has a string spectrum that contains both open and closed string states with a (string-)Galilean-invariant dispersion relation.
The BRST structure NR string theory is the same as in relativistic string theory, so its critical dimensions are $d=26$ for bosonic string theory and $d=10$ for superstring theories~\cite{Gomis:2000bd}.
Intriguingly, in order for NR string theory to have a nonempty string spectrum that contains propagating degrees of freedom, it turns out that we have to compactify the longitudinal spatial direction $X^1$ over a circle, as we will see in the following.

\subsection{Nonrelativistic string theory as a low-energy limit} \label{sec:zeroRegge}

Although NR string theory can be studied from first principles using the action \eqref{eq:cg}, it is useful to understand how this theory is embedded in relativistic string theory.
In fact, historically, NR string theory was initially introduced as a zero Regge slope limit of relativistic string theory in a near-critical $B$-field \cite{Klebanov:2000pp, Gomis:2000bd, Danielsson:2000gi}.
Our starting point is the sigma model that describes relativistic string theory,
\begin{equation} \label{eq:Srel}
    \hat{S} = \frac{1}{4\pi\hat{\alpha}'} \int_\Sigma d^2 \sigma \, \Bigl( \p_\alpha X^\mu \, \p^\alpha X^\nu \, \hat{G}_{\mu\nu} - i \, \epsilon^{\alpha\beta} \, \p_\alpha X^\mu \, \p_\beta X^\nu \, \hat{B}_{\mu\nu} \Bigr)\,,
\end{equation}
with the following Riemannian (or Lorentzian) metric and Kalb-Ramond background fields:
\begin{align} \label{eq:ncoslimit}
    \hat{G}_{\mu\nu} =
        \begin{pmatrix}
            \eta^{}_{AB} &\,\, 0 \\[4pt]
            0 &\,\, \frac{\hat{\alpha}'}{\alpha'} \, \delta_{A'B'}
        \end{pmatrix}\,,
        \qquad%
    \hat{B}_{\mu\nu} =
        \begin{pmatrix}
            - \epsilon^{}_{AB} &\,\, 0 \\[4pt]
            0 &\,\, 0
        \end{pmatrix}\,.
\end{align}
Here and in the following, we use hats to distinguish variables in relativistic string theory, while variables in NR string theory are unhatted.
On this background, the relativistic string action~\eqref{eq:Srel} is
\begin{align} \label{eq:rel2}
    \hat{S} = \frac{1}{4\pi\alpha'} \int_\Sigma d^2 \sigma \, \bigg( \p_\alpha X^{A'} \p^\alpha X_{A'} - \frac{\alpha'}{\hat{\alpha}'} \, \bar{\p} X \, \p \overline{X} \biggr)\,.
\end{align}
This action seems to  be singular in the $\hat{\alpha}' \to 0$ limit.
To obtain a finite action under this limit, we introduce
\begin{equation} \label{eq:Sreleq}
    \hat{S} = \frac{1}{4\pi\alpha'} \int_\Sigma d^2 \sigma \, \bigg( \p_\alpha X^{A'} \p^\alpha X_{A'} + \lambda \, \bar{\p} X + \bar{\lambda} \, \p \overline{X} + \frac{\hat{\alpha}'}{\alpha'} \, \lambda \bar{\lambda} \biggr)\,,
\end{equation}
which reproduces the action~\eqref{eq:rel2} upon integrating out the auxiliary fields $\lambda$ and $\bar{\lambda}$ in the path integral.
Taking the limit $\hat{\alpha}' \rightarrow 0$ in \eqref{eq:Sreleq} gives rise to a finite action, which is the same as the NR string action~\eqref{eq:cg}, with $\alpha'$ being the effective Regge slope in NR string theory.
The associated interactions are only finite if we simultaneously send the relativistic string coupling $\hat{g}_s$ to infinity, while holding $\hat{g}_s \, \hat{\alpha}{}'{}^{1/2}$ fixed (which corresponds to the radius of the circle compactified over the eleventh dimension in M-theory).
Under this double scaling limit, the resulting NR string theory has an effective string coupling $g_s = \hat{g}_s \sqrt{\hat{\alpha}{}' / \alpha'}$\,, where both $g_s$ and the effective Regge slope $\alpha'$ are finite.
This limit\footnote{%
    Although this limit is the main focus of this review, several other limits of the relativistic string~\eqref{eq:Sfree} can be considered.
    For example, a different NR limit of relativistic string theory has been explored in~\cite{Batlle:2016iel}, where only the time direction instead of the two-dimensional longitudinal sector is treated differently.
    This limit leads to NR strings that do not vibrate.
    Additionally, a tensionless limit of relativistic string theory has been considered~\cite{Schild:1976vq,Isberg:1993av,Bagchi:2013bga}, which leads to a sigma model with non-Riemannian worldsheet structure, similar to the further limit of NR strings we will discuss in \S\ref{sec:nhd}.
}
is also known as the noncommutative open string (NCOS) limit~\cite{Seiberg:2000ms, Gopakumar:2000na, Klebanov:2000pp}.
We will discuss its connection to NCOS in \S\ref{sec:nrosncos}.

We now examine the closed string states. The constant $B$-field in \eqref{eq:Srel} has a nontrivial effect if the $X^1$ direction is compactified over a circle of radius $R$\,.
In the relativistic string theory described by~\eqref{eq:Srel}, closed string states with a nonzero winding number $w$ in $X^1$ and momentum $K_\mu$ satisfy the following mass-shell condition (see for example \cite{Grignani:2001hb}):
\begin{equation} \label{eq:massshell}
    \biggl( E + \frac{w R}{{\hat{\alpha}}'} \biggr)^{\!2} - \frac{\alpha'}{\hat{\alpha}'} \, K^{A'} K_{A'} = \frac{n^2}{R^2} + \frac{w^2 R^2}{{\hat{\alpha}}'{}^2} + \frac{2}{{\hat{\alpha}}'} \, (N + \tilde{N} - 2)\,,
\end{equation}
together with the level-matching condition $n \, w = \tilde{N} - N$.
Here, $n$ is the Kaluza-Klein number and $(N, \tilde{N})$ are the string excitation numbers. The shift of the energy $E$ in \eqref{eq:massshell} is due to the constant $B$-field in the compactified $X^1$ direction. In the $\hat{\alpha}' \rightarrow 0$ limit, we find the dispersion relation for NR closed strings,
\begin{equation}
  \label{eq:dr}
  E = \frac{\alpha'}{2 w R} \biggl[ K^{A'} K_{A'} + \frac{2}{\alpha'} \, (N + \tilde{N} - 2) \biggr]\,.
\end{equation}
Finiteness of the dispersion relation \eqref{eq:dr} imposes the condition that $w \neq 0$\,.
Therefore, all asymptotic states in the closed string spectrum necessarily carry a nonzero string winding number along the compact $X^1$ direction \cite{Gomis:2000bd, Danielsson:2000mu}.
Note that the Kaluza-Klein momentum number~$n$ does not show up explicitly in the dispersion relation, but only enters via the level-matching condition.

As is evident from the rewriting \eqref{eq:rel2} of relativistic string theory, the free theory \eqref{eq:cg} that describes NR string theory can be deformed towards relativistic string theory by reintroducing the operator $\lambda\bar{\lambda}$ as in the action~\eqref{eq:Sreleq}, with a nonzero $\hat{\alpha}'$ \cite{Danielsson:2000mu}. Indeed, turning on the $\lambda\bar{\lambda}$ deformation controlled by the coupling $U_0 = \hat{\alpha}' / \alpha'$ inside the NR string action~\eqref{eq:cg} modifies the NR dispersion relation \eqref{eq:dr} back to \eqref{eq:massshell}, which we rewrite as
\begin{equation}
  \label{eq:dispersion-including-lambda-lambdabar}
  E = \frac{\alpha'}{2 w R} \left[ K^{A'} K_{A'} + \frac{2}{\alpha'} \bigl( N + \tilde{N} - 2 \bigr) - U_0 \left( E^2 - \frac{n^2}{R^2} \right) \right].
\end{equation}
When $U_0 \neq 0$\,, there are asymptotic states in the zero-winding sector with $w=0$ that satisfy the relativistic dispersion relation,
\begin{equation}
  \label{eq:reldr}
  U_0 \, \left( E^2 - \frac{n^2}{R^2} \right) - K^{A'} K_{A'} = \frac{2}{\alpha'} \bigl( N + \tilde{N} - 2 \bigr)\,.
\end{equation}
In contrast, as we have seen earlier, only states corresponding to strings that have nonzero winding around the longitudinal target space circle $X^1$ survive in the NR string theory limit $\hat{\alpha}'\to0$\,.
To identify our NR corner in string theory, the $\lambda\bar{\lambda}$ deformation that drives the theory away from the NR regime must therefore be eliminated. We will review how the $\lambda\bar{\lambda}$~deformation is treated in the literature later in \S\ref{sec:nscs}, where string interactions are included.

\subsection{Nonrelativistic closed strings}
\label{ssec:flat-nr-closed-strings}

Having reviewed how NR string theory arises as a zero Regge slope limit in string theory, we return to the defining action \eqref{eq:cg} for NR string theory, focusing on the sector of \emph{nonrelativistic closed string} (NRCS) theory.
We already learned from the $\alpha'\to0$ limit that, in order to have a nonempty closed string spectrum, we have to compactify the longitudinal spatial direction~$X^1$ over a spatial circle of radius $R$.
We will now see that the Galilean-invariant dispersion relation \eqref{eq:dr} can be derived directly from the NR string action~\eqref{eq:cg}, without performing any limits, by constructing the BRST-invariant vertex operators in NRCS.
For this, we first discuss a physical interpretation of the $\lambda$ and $\bar\lambda$ fields by considering T-duality transformations of NRCS.

We first consider a T-duality transformation along the compact longitudinal target space direction $X^1$.
For this, we introduce the parent action
\begin{equation}
    \label{eq:flat-t-duality-first-aux-action}
    S = \frac{1}{4\pi \alpha'} \int d^2\sigma
    \left[
        \pd_\alpha X^{A'} \pd^\alpha X_{A'}
        + \lambda \left( \bar\pd X^0 + \bar v \right)
        + \bar\lambda \left( \pd X^0 - v \right)
        + 2 Y_1 \left( \bar\pd v - \pd \bar v \right)
    \right].
\end{equation}
Integrating out $Y_1$ imposes $\bar\pd v = \pd \bar v$, which we can solve locally by setting $v = \pd X^1$ and $\bar v = \bar\pd X^1$.
This reproduces the conformal gauge NR string action~\eqref{eq:cg}.
Instead, we integrate out $v$ and $\bar v$\,, which imposes $\lambda = - 2 \, \pd Y_1$ and $\bar\lambda = -2 \, \bar\pd Y_1$, so the action~\eqref{eq:flat-t-duality-first-aux-action} reduces to
\begin{equation}
    \label{eq:flat-dlcq-action}
    S = \frac{1}{4\pi \alpha'} \int d^2\sigma
    \left(
        \pd_\alpha X^{A'} \pd^\alpha X_{A'}
        - 2 \pd Y_1 \bar\pd X^0
        - 2 \bar\pd Y_1 \pd X^0
    \right).
\end{equation}
This is the action of relativistic string theory in a flat background, with spatial directions~$X^{A'}$ and lightlike directions $X^0$ and $Y_1$.
However, since it is dual to $X^1$, which is a circle with radius~$R$, the lightlike direction $Y^1$ is a circle with effective radius $\alpha'/R$.
Therefore, the closed string described by the action~\eqref{eq:flat-dlcq-action} describes the DLCQ of relativistic string theory~\cite{Gomis:2000bd, Danielsson:2000mu}.
As such, NRCS provides a NR covariant definition of DLCQ in relativistic string theory, which is normally defined using a subtle limit of the compactification of relativistic strings on a spacelike circle \cite{Seiberg:1997ad, Sen:1997we, Hellerman:1997yu}.

\begin{figure}
  \centering
  \begin{tikzpicture}
    \node[draw, rectangle, minimum width = 2cm, minimum height = 1cm, align = center]
      (UL) at (0,4)
      {%
        NRCS
      };
    \node[draw, rectangle, minimum width = 2cm, minimum height = 1cm, align = center]
      (UR) at (8,4)
      {%
        NRCS
      };
    \node[draw, rectangle, minimum width = 2cm, minimum height = 1cm, align = center]
      (DL) at (0,0)
      {%
        DLCQ
        \\
        of rel.\ string
      };
    \node[draw, rectangle, minimum width = 2cm, minimum height = 1cm, align = center]
      (DR) at (8,0)
      {%
        rel.\ strings
      };

    \draw [-Stealth] (UL) -- node [above] {%
        transverse
        } node [below] {%
        T-duality
      }
      (UR);
    \draw [-Stealth] (UR) -- (UL);
    \draw [-Stealth] (DR) -- node [above] {%
        lightlike
        } node [below, align = center] {%
        compactification
      }
      (DL);
    \draw [-Stealth] (UL) -- node [left] {%
        } node [right, align = left] {%
        longitudinal spatial
        \\
        T-duality
      }
      (DL);
  \end{tikzpicture}
  \caption{T-duality transformations of NR closed string theory (NRCS).}
  \label{fig:nrcs-diagram}
\end{figure}
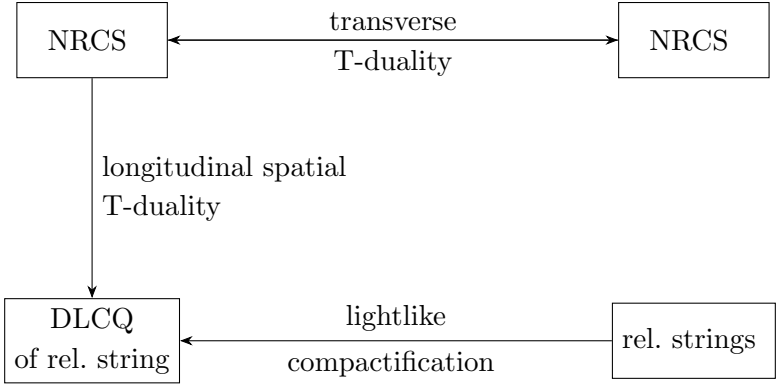

A T-duality transformation of NRCS along a compact transverse direction acts in the same way as in relativistic string theory, resulting in NRCS on a background with the corresponding dual compact transverse direction.
The complete curved-spacetime Buscher rules can be found in \cite{Bergshoeff:2018yvt, Bergshoeff:2019pij, Kluson:2018vfd}, see also \cite{Harmark:2017rpg, Kluson:2018egd, Harmark:2018cdl, Harmark:2019upf, Kluson:2019xuo} for related works.
These T-duality relations of NRCS are displayed in Figure~\ref{fig:nrcs-diagram}.

In addition, we consider a T-duality transformation along a lightlike longitudinal direction in the action \eqref{eq:flat-dlcq-action} that describes the DLCQ of relativistic string theory.
To do this, we Wick rotate and compactify the $X^0$ direction in the original action \eqref{eq:cg}.
Following the same procedure as before, we start from the action~\eqref{eq:flat-dlcq-action} and exchange the $X^0$ direction for a dual~$Y_0$, which leads to
\begin{equation}
    S = \frac{1}{4\pi \alpha'} \int d^2\sigma
    \left(
        \pd_\alpha X^{A'} \pd^\alpha X_{A'}
        + u \, \bar\pd Y
        - \bar u \, \pd \bar Y
    \right),
\end{equation}
where $Y = Y_0 + Y_1$ and $\bar Y = Y_0 - Y_1$.
This action describes NRCS.
Additionally, since $u$ and~$\bar u$ impose the constraints $\pd \bar Y= 0$ and $\bar\pd Y = 0$, we get the duality map
\begin{equation} \label{eq:Tduallambda}
    \lambda = -2 \, \pd Y_1 = - \pd Y,
    \qquad
    \bar\lambda = - 2 \, \bar \pd Y_1 = \bar\pd \bar Y.
\end{equation}
These equations map the NRCS one-forms $\lambda$ and $\bar\lambda$ to the dual coordinates $Y$ and $\bar Y$\,.
As such, we can interpret the one-forms on the worldsheet as conjugates to the longitudinal string winding, whereas the $X^A$ coordinates are conjugate to the longitudinal string momentum~\cite{Gomis:2020fui,Yan:2021lbe}.
This can also be understood from a double field theory perspective~\cite{Ko:2015rha,Morand:2017fnv}.

While the relation \eqref{eq:Tduallambda} is technically only valid for compact $X^0$, we can still use $\lambda = - \pd Y$ and $\bar \lambda = \bar\pd Y$ (together with $\bar\pd Y = 0$ and $\pd \bar Y = 0$) as a field redefinition~\footnote{The field redefinition \eqref{eq:Tduallambda} involves time derivatives and contributes nontrivially to the path-integral measure. However, in the operator formalism, the substitution \eqref{eq:Tduallambda} is always valid.} to obtain a convenient set of of parameters for the operator product expansions (OPEs) of the original NRCS string action~\eqref{eq:cg}.
In radial quantization, we define $z = e^{\sigma^2 + i \sigma^1}$ and $\bar{z} = e^{\sigma^2 - i \sigma^1}$.
In terms of $X^{A'}$, $X^A$ and the dual variables $Y_A$, the nontrivial OPEs are~\footnote{A further reparametrization of worldsheet fields that mix $X^A$ and $Y_A$ has been considered in \cite{Yan:2021lbe}, where the resulting OPEs take the same form as in relativistic string theory. It is therefore possible to evaluate string amplitudes in NR string theory by borrowing results directly from relativistic string theory.}
\begin{align}
\begin{split}
    :\! Y (z) \, X (z') \!: \, & \sim \alpha' \ln \bigl( z - z' \bigr)\,,
        \qquad%
    :\! X^{A'} (z) \, X^{B'} (z') \!: \, \sim - \frac{1}{2}\, \alpha'\,  \delta^{A'B'} \ln \left| z - z' \right|^2\,, \\[2pt]
    :\! \overline{Y} (\bar{z})\, \overline{X} (\bar{z}') \!: \, & \sim - \alpha' \ln \bigl( \bar{z} - \bar{z}' \bigr)\,.
\end{split}
\end{align}
The closed string tachyon vertex operator then takes the form
\begin{equation}
  \label{eq:tvo}
  :\! e^{i \, ( K_\mu \, X^\mu + \, Q^A \, Y_A )} \!:\,,
\end{equation}
where $Q^0 = 0$ and $Q^1 = - 2 w R / \alpha'$ parametrizes the longitudinal winding.
We have omitted a cocycle factor, which is needed for the single-valuedness of the OPEs and contributes a sign to string amplitudes.
Higher-order vertex operators are constructed from the tachyon vertex operator \eqref{eq:tvo} by dressing it up with derivatives of $X^\mu$ and $Y_A$\,. The BRST invariance of such vertex operators then leads to the dispersion relation \eqref{eq:dr} \cite{Gomis:2000bd, Yan:2021lbe}.

The string amplitudes between winding closed strings represented by such vertex operators have been considered in \cite{Gomis:2000bd}.
The tree-level string amplitudes have poles corresponding to excited closed string states carrying nonzero winding.
There is no graviton in the spectrum of NRCS. However, in the special case where the winding number is not exchanged among the asymptotic states, the amplitudes
gains a contribution from exchanging \emph{off-shell} states in the zero winding sector. The leading long-range contribution is proportional to $1/ (K_{A'} K^{A'})$\,. These zero-winding states
become of measure zero in the asymptotic limit, and therefore only arise as intermediate states \cite{Danielsson:2000mu}. These intermediate states give rise to a Newton-like potential after a Fourier transform, and induce an instantaneous gravitational force between winding strings.

As in relativistic string theory, NRCS has a perturbative expansion with respect to the genera of the worldsheet Riemann surfaces. However, at loop level, there are nontrivial constraints that restrict the moduli space to a lower dimensional manifold~\cite{Gomis:2000bd}.
This is because the one-form fields ($\lambda$, $\bar{\lambda}$) play the role of Lagrange multipliers that require ($X$,$\bar{X}$) to be (anti-)holomorphic maps in~\eqref{eq:holocond} from the worldsheet to the longitudinal sector of the target space.
For example, the bosonic one-loop free energy at the inverse temperature $\beta$ has been analyzed in~\cite{Gomis:2000bd}.
This free energy determines the thermodynamic partition function of free closed strings and gives rise to the Hagedorn temperature.
It requires a Wick rotation of~$X^0$ in the target space, followed by a periodic identification $X^0 \sim X^0 + \beta$\,. The path integral over the zero modes of $\lambda$ and $\bar{\lambda}$ leads to the following constraint on the worldsheet modulus~$\tau$\,:
\begin{equation}
  \label{eq:mconst}
  \tau = \frac{1}{w} \left( n + \frac{i m \beta}{2\pi R} \right),
  \qquad
  n, \, m\,, w\, \in \, \mathbb{Z}\,.
\end{equation}
Here, $m$ denotes the winding number in $X^0$. From \eqref{eq:mconst}, it is manifest that the integral over the fundamental domain for the moduli space of the torus in the evaluation of one-loop amplitudes is now localized to be a sum over discrete points.
{The fact that the one-loop moduli space for NR strings lies within the fundamental domain for relativistic string theory, implies that the NR string free energy is finite.}
The constraint \eqref{eq:mconst} is also generalized to higher-loop and general $N$-point amplitudes \cite{Gomis:2000bd}, in such a way that holomorphic maps from the worldsheet to the target space exist. Such localization theorems in the moduli space suggest that the computation of NR string amplitudes may simplify significantly compared to the case in relativistic string theory.~\footnote{See~\cite{Yan:2021hte} for generalizations of such localization theorems at one-loop to NR open strings. In this paper, KLT relations between tree-level NR string amplitudes are also studied.} The free energy and $N$-point amplitudes at one-loop order match the ones in the DLCQ of string theory \cite{Bilal:1998vq, Danielsson:2000mu}.~\footnote{It is also shown in \cite{Grignani:1999sp} that the thermodynamic partition function of the finite temperature type IIA string theory in the DLCQ is equivalent to the partition function of matrix string theory.}

\subsection{Nonrelativistic and noncommutative open strings} \label{sec:nrosncos}

We now consider open strings, whose worldsheet $\Sigma$ has a boundary $\p \Sigma$\,.
At tree level, $\Sigma$~is a strip with $\sigma^1 \in [0,\pi]$ and the Euclidean time $\sigma^2 \in \mathbb{R}$\,.
Depending on which boundary conditions the open strings satisfy in the compactified $X^1$ direction, there are two open string sectors that are associated to the defining action \eqref{eq:cg}: (i) the \emph{nonrelativistic open string} (NROS) sector with NR string spectrum that has a Galilean-invariant dispersion relation~\cite{Danielsson:2000mu}, and (ii) the \emph{noncommutative open string} (NCOS) sector with noncommutativity between space and time~\footnote{This space/time noncommutativity is tied to the stringy nature of the theory. In contrast, introducing noncommutativity between space and time in field theories typically leads to inconsistencies \cite{Seiberg:2000gc, Barbon:2000sg, Gomis:2000zz}. Also see \cite{Aharony:2000gz} for theories with lightlike noncommutativity.} and a relativistic string spectrum \cite{Gomis:2000bd, Danielsson:2000gi, Danielsson:2000mu}.
For simplicity, we require in the following discussions that the open strings satisfy Neumann boundary conditions in $X^0$ and~$X^{A'}$, with
$\p X^0 / \p \sigma^1 = \p X^{A'} / \p \sigma^1 = 0$ on $\p\Sigma$ at $\sigma^1 = 0, \pi$\,.

First, consider open strings that satisfy the Dirichlet boundary condition $\delta X^1 = 0$ on~$\p\Sigma$\,,
by anchoring the ends of an open string on D$(d-2)$-branes transverse to $X^1$. In this case, the variation of the action \eqref{eq:cg} with respect to $X^A$ vanishes only if
$\lambda = \bar{\lambda}$ on $\p \Sigma$\,. The open string spectrum has a Galilean-invariant dispersion relation~\cite{Danielsson:2000mu},
\begin{equation}
  K_0 = \frac{\alpha'}{2wR} \left[ K^{A'} K_{A'} + \frac{1}{\alpha'} (N-1) \right],
\end{equation}
where $w$ is the fractional winding number of open strings stretched between transverse D-branes located along $X^1$.
Therefore, imposing Dirichlet boundary conditions in $X^1$ defines the NROS sector. On the D-brane, the global symmetry of the sigma model is broken down to be the Bargmann symmetry. In the zero winding sector, the effective field theory living on a stack of $n$ coinciding D-branes is Galilean Yang--Mills theory \cite{Gomis:2020fui},
\begin{equation}
  \label{eq:gym}
  S_\text{YM} = \frac{1}{g_{\text{YM}}^2} \int dX^0 \, dX^{A'} \, \text{tr} \Bigl( \tfrac{1}{2} \, D_0 N \, D_0 N - E_{A'} \, D^{A'} N - \tfrac{1}{4} \, F_{A'B'} \, F^{A'B'} \Bigr)\,,
\end{equation}
with $D_0$ and $D_{A'}$ are covariant derivatives with respect to the $U(n)$ gauge group. The electric and magnetic field strengths $E_{A'}$ and $F_{A'B'}$ are associated to the gauge fields $A_0$ and $A_{A'}$ on the D-brane. The scalar field $N$ is in the adjoint representation of $U(n)$\,, and perturbs around the solitonic D$(d-2)$-brane. In the $U(1)$ case, this gives rise to Galilean electrodynamics (GED) \cite{le1973galilean, Santos:2004pq, Festuccia:2016caf}.~\footnote{%
Note that this theory contains no propagating degrees of freedom.
However, in \cite{Chapman:2020vtn}, it is shown that coupling GED to Schr\"{o}dinger scalars in $2+1$ dimensions affects the renormalization group (RG) structure nontrivially and leads to a family of NR conformal fixed points.
}

Next, consider open strings that satisfy the Neumann boundary condition $\p X^1 / \p \sigma^1 = 0$ on $\p \Sigma$\,. In this case, open strings reside on spacetime filling D-branes. For the theory to be well-defined, a nonzero electric field strength $E$ (or a nonzero $B$-field) is introduced.
The resulting theory has a relativistic string spectrum
and noncommutativity between the longitudinal space and time directions, with $\bigl[ X^0, X^1 \bigr] \propto E^{-1}$\,.
Therefore, imposing Neumann boundary conditions in $X^1$ defines the NCOS sector \cite{Gomis:2000bd, Danielsson:2000gi}.
NCOS was first discovered as a low energy limit of string theory \cite{Seiberg:2000ms, Gopakumar:2000na, Klebanov:2000pp}, in the same setup that we discussed in \S\ref{sec:zeroRegge}.
Also see \cite{Connes:1997cr, Douglas:1997fm, Seiberg:1999vs} for original works on D-branes in magnetic fields and their applications to noncommutative Yang--Mills theories.
NCOS is S-dual to spatially-noncommutative Yang--Mills theory \cite{Gopakumar:2000na}.

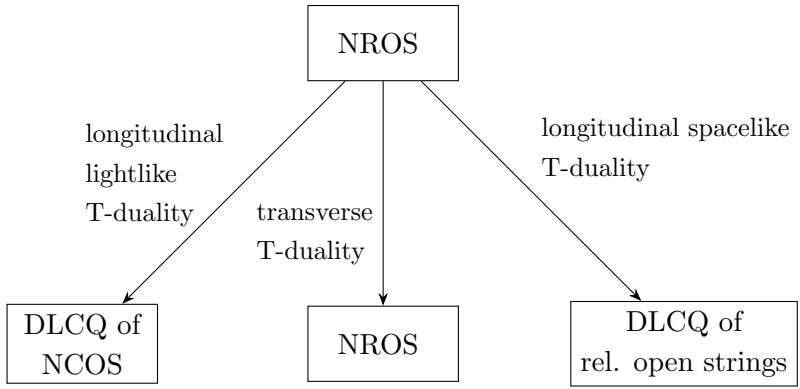
\begin{figure}
  \centering
  \begin{tikzpicture}
    \node[draw, rectangle, minimum width = 2cm, minimum height = 1cm, align = center]
      (UL) at (4,4)
      {%
        NROS
      };
    \node[draw, rectangle, minimum width = 2cm, minimum height = 1cm, align = center]
      (UR) at (4,0)
      {%
        NROS
      };
    \node[draw, rectangle, minimum width = 2cm, minimum height = 1cm, align = center]
      (DL) at (0,0)
      {%
        DLCQ of
        \\
        NCOS
      };
    \node[draw, rectangle, minimum width = 2cm, minimum height = 1cm, align = center]
      (DR) at (8,0)
      {%
        DLCQ of\\
        rel. open strings
      };

    \draw [-Stealth] (UL) -- node [below left, align = left] {%
        \small
        transverse
        \\
        \small
        T-duality
      }
      (UR);
    \draw [-Stealth] (UL) -- node [left] {%
        } node [left, align = left] {%
        \small
        longitudinal
        \\
        \small
        lightlike
        \\
        \small
        T-duality\\\mbox{}
      }
      (DL);
    \draw [-Stealth] (UL) -- node [left] {%
        } node [above right, align = left] {%
        \small
        longitudinal spacelike
        \\
        \small
        T-duality
      }
      (DR);
  \end{tikzpicture}
  \caption{T-duality relations for NR open string theory (NROS).
    The transverse T-duality swaps Dirichlet and Neumann boundary conditions in the usual way.
  }
  \label{fig:nros-diagram}
\end{figure}

Nonrelativistic and noncommutative open strings are related via T-duality \cite{Gomis:2020izd}, as illustrated in Figure~\ref{fig:nros-diagram}.
In NROS, the geometry of the longitudinal sector in the target space is taken to be a spacetime cylinder, wrapping around the compactified longitudinal spatial direction~$X^1$.
Performing a T-duality transformation along $X^1$ in NROS leads to the DLCQ of relativistic open string theory on spacetime filling D-branes. To make the connection to NCOS, one needs to introduce a twist in the compactification of $X^1$ by shifting one end of the longitudinal cylinder along the time direction, before gluing back. This shift does not change the nature of the T-duality transformation and still leads to the DLCQ of relativistic open strings, unless the shift equals the circumference of the longitudinal circle. In the latter case, the T-dual theory is NCOS on a spacetime-filling brane and with a compact longitudinal lightlike circle. The background electric field in NCOS corresponds to a rescaling factor of the $X^1$ circle in NROS. It is also interesting to consider a T-duality transformation along~$X^1$ in NCOS. In the T-dual frame, there arises relativistic open string theory on a D$(d-2)$-brane in the DLCQ description. Such a D$(d-2)$-brane is infinitely boosted along a spatial circle~\cite{Seiberg:2000ms}. Generalizations of the above T-duality transformations in arbitrary background fields are studied in \cite{Gomis:2020izd}.

\section{Nonrelativistic Strings in Curved Spacetime} \label{sec:nscs}

After reviewing the basic ingredients of NR string theory in flat spacetime, we now consider generalizations to curved spacetime.
In the following, we will restrict to string states with zero winding along the compact longitudinal direction.
Such states are not part of the physical spectrum, but they serve to mediate the instantaneous forces between the physical asymptotic states with nonzero winding.
As a result, the low-energy effective theory that arises from the nonwinding sectors of closed and open strings that we consider in the following play a similar role to the instantaneous force in Newtonian gravity or the Coulomb force in electrostatics.
Exponentiating the vertex operators associated with such zero winding states in the path integral gives rise to various background fields. These background fields are functional couplings in the nonlinear sigma model that generalizes the free worldsheet theory \eqref{eq:Sfree} by including arbitrary marginal deformations that are conformally invariant.

We have seen in \S\ref{sec:zeroRegge} that the marginal operator $\lambda \bar{\lambda}$ drives the theory towards the relativistic regime. In particular, this operator deforms the NR dispersion relation \eqref{eq:dr} to the relativistic dispersion relation \eqref{eq:reldr}.
In this sense, the free action \eqref{eq:Sfree}
defines an unconventional vacuum around which string theory can be expanded.
As shown in \S\ref{sec:zeroRegge}, NR string theory is defined at the corner where the theory is tuned such that no $\lambda \bar{\lambda}$ counterterms are generated.
In the following, we start by considering NR string sigma models
where the $\lambda\bar{\lambda}$ operator on the worldsheet is classically tuned to be zero.
The consequences of this tuning at the quantum level will be discussed in \S\ref{sec:eftns}.

The remaining background fields give rise to a general framework for studying the appropriate spacetime geometry coupled to NR string theory.
The resulting target space geometry is known as \emph{torsional string Newton-Cartan} (TSNC) geometry~\cite{Bidussi:2021ujm,Harmark:2019upf, Bergshoeff:2021bmc, Gallegos:2020egk}, since it generically allows for nonzero torsion.
In contrast to Newton--Cartan geometry, which is related to particle probes, string Newton--Cartan geometry contains not one but two distinguished directions that are longitudinal to the string.
In the free worldsheet action~\eqref{eq:Sfree}, these directions are represented by the longitudinal lightlike coordinates $X$ and~$\overline{X}$.
We discuss the gauge symmetries associated to the TSNC target space geometry and show how they can be obtained by gauging a Lie algebra.
We also illustrate the connection to double field theory and null reduction.

\subsection{Strings in torsional string Newton--Cartan geometry} \label{sec:smbf}

The curved spacetime generalization of the free NR string theory action~\eqref{eq:Sfree} is obtained by turning on all allowed marginal local interactions in the sigma model, which leads to the classically-conformal action \cite{Gomis:2005pg, Bergshoeff:2018yvt}
(see also \cite{Yan:2021lbe} for the inclusion of the $\lambda\bar\lambda$ term)
\begin{align} \label{eq:cbsa}
\begin{split}
    \hat{S} = \frac{1}{4\pi\alpha'} \int_\Sigma d^2 \sigma \, \Bigl\{ & \sqrt{h} \, h^{\alpha\beta} \, \p_\alpha X^{\mu} \, \p_\beta X^{\nu} \, S_{\mu\nu}(X) - i \, \epsilon^{\alpha\beta} \, \p_\alpha X^\mu \, \p_\beta X^\nu \, A_{\mu\nu}(X) \\[2pt]
    & \!\!\!\!\!\!\!\!\!\!\! + \sqrt{h} \, \left[ \lambda \, \bar{\mathcal{D}} X^\mu \, \tau_\mu(X) + \bar{\lambda} \, \mathcal{D} X^\mu \, \bar{\tau}_\mu(X) + \lambda \bar{\lambda} \, U(X) + \alpha' \, \text{R} (h) \, \Phi(X) \right] \Bigr\}\,.
\end{split}
\end{align}
Here, $h_{\alpha\beta}$ is the worldsheet metric, R$(h)$ is its Ricci scalar, and $\lambda$ and $\bar\lambda$ are one-forms on the worldsheet.
The symmetries of the sigma model consist of the standard worldsheet diffeomorphisms, worldsheet Weyl invariance, and target space reparametrizations.
If $U\neq0$\,, the one-form fields can be integrated out, and we end up with relativistic string theory.
In the following, we first discuss the geometry associated to the classical NR string theory at $U=0$\,, and we return to
the interplay between the $U\to0$ limit and quantum effects in \S\ref{sec:eftns}.

The background fields in this action consist of the symmetric and antisymmetric two-tensors $S_{\mu\nu}$ and $A_{\mu\nu}$, the one-forms $\tau_\mu$ and $\bar\tau_\mu$ and the dilaton $\Phi$.
They can be interpreted as a coherent state of NR strings.
Demanding that~\eqref{eq:cbsa} is invariant under reparametrizations of $X^\mu$ implies that the background fields transform covariantly under general target-space diffeomorphisms.
Furthermore, we introduce coordinates $x^{\hmu}$ that form a chart of the curved target-space manifold.
As a result, the worldsheet fields $X^\mu$ are the composition of
$x^\mu$ and the embedding of the worldsheet in the target space.
Note that we only consider marginal couplings, and we only allow the background fields to depend on the embedding fields $X^\mu$, which include both the longitudinal and transverse directions.
We do not allow the background fields to depend on the one-forms $\lambda$ and $\bar\lambda$, which are associated with vertex operators that correspond to winding string states.

One of the remarkable features of the resulting target-space geometry is that it contains the one-forms $\tau_\mu{}^A$, where $A=0,1$, which can be interpreted as vielbeine that parametrize the directions that are longitudinal to the string.
These longitudinal vielbeine come with a corresponding Minkowski metric $\eta^{}_{AB}$ and they are related to the fields in the action~\eqref{eq:cbsa} by
\begin{equation}
  \tau_\mu{}^0 = \frac{1}{2} \left(\tau_\mu + \bar\tau_\mu\right),
    \qquad%
  \tau_\mu{}^1 = \frac{1}{2} \left(\tau_\mu - \bar\tau_\mu\right).
\end{equation}
In addition, the worldsheet couplings contain the symmetric and antisymmetric two-tensors $S_{\mu\nu}$ and $A_{\mu\nu}$.
However, the action~\eqref{eq:cbsa} is invariant under a set of St\"{u}ckelberg-type transformations \cite{Bergshoeff:2018yvt},
\begin{align} \label{eq:Stueckelberg}
    S_{\mu\nu} \rightarrow S_{\mu\nu} - 2 \, \mathcal{C}_{(\mu}{}^A \, \tau_{\nu)}{}^B \, \eta_{AB}\,,
        \qquad%
    A_{\mu\nu} \rightarrow A_{\mu\nu} + 2 \, \mathcal{C}_{[\mu}{}^A \, \tau_{\nu]}{}^B \, \epsilon_{AB}\,,
\end{align}
together with appropriate shifts of the Lagrange multipliers $\lambda$ and $\bar\lambda$ that impose the constraints involving the longitudinal vielbeine.
Here, $\epsilon_{AB}$ is the Levi-Civita symbol for the longitudinal directions, and $\mC_\mu{}^A$ is an arbitrary matrix.
This St\"{u}ckelberg symmetry~\eqref{eq:Stueckelberg} allows one to shuffle the geometric degrees of freedom in the longitudinal directions between the symmetric and antisymmetric couplings $S_{\mu\nu}$ and $A_{\mu\nu}$.

\begin{figure}[t]
\begin{minipage}{6in}
  \centering
  \includegraphics[align=c,height=2in]{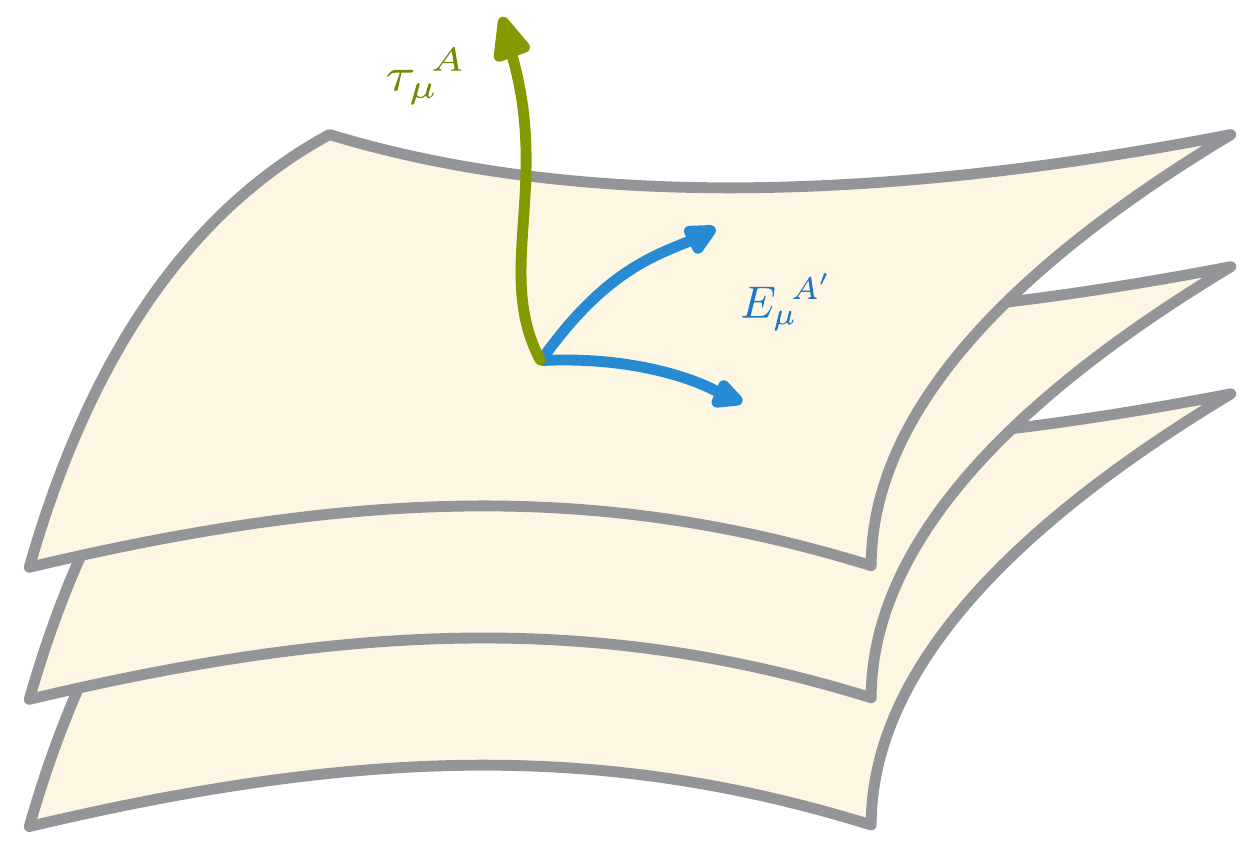}
  \hspace*{.2in}
  \includegraphics[align=c,height=2in]{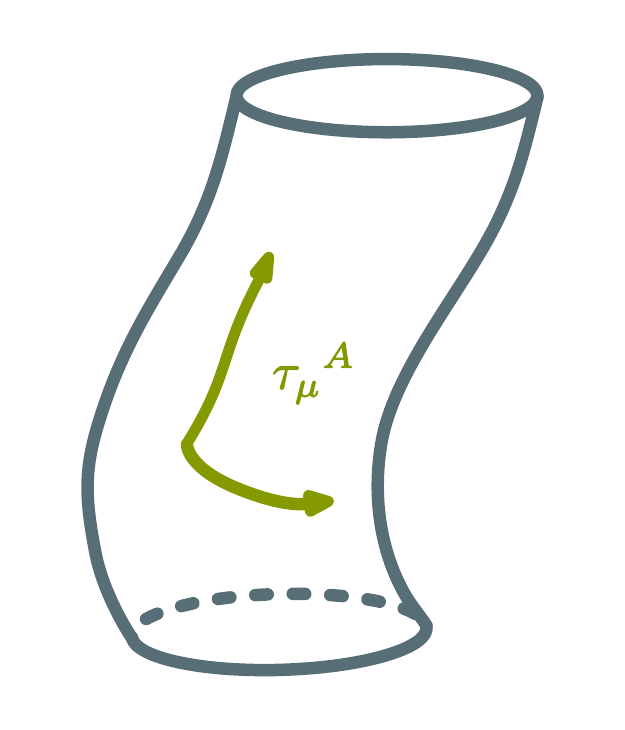}
\end{minipage}
    \caption{A schematic illustration of a stringy generalization of Newton--Cartan geometry.
    The green arrows represent the two longitudinal directions along the worldsheet, whose geometry is encoded in the vielbein fields $\tau_\mu{}^A$. Each horizontal slice represents the $(d-2)$-dimensional transverse directions, parametrized by the vielbein fields $E_\mu{}^{A'}$.}
    \label{fig:bread}
\end{figure}

We can fix this St\"{u}ckelberg symmetry by requiring $S_{\mu\nu}$ to be fully transverse with respect to the longitudinal directions.
For this, we introduce a set of inverse vielbeine $\tau^\mu{}_A$ such that $\tau_\mu{}^A \, \tau^\mu{}_B = \delta^A_B$\,, and set $\tau^\mu{}_A \, S_{\mu\nu} \to 0$\,.
We denote the resulting couplings by \cite{Harmark:2019upf}
\begin{equation}
\label{eq:stueckelberg-fix-no-long-metric}
  S_{\mu\nu} \to E_{\mu\nu} = \delta_{A'B'} \, E_\mu{}^{A'} E_\nu{}^{B'},
    \qquad%
  A_{\mu\nu} \to M_{\mu\nu}\,.
\end{equation}
Here, $M_{\mu\nu}$ is still an arbitrary antisymmetric tensor, which generically contains both transverse and longitudinal components, but $E_{\mu\nu}$ is now purely transverse.
For this reason, we have introduced the transverse vielbeine $E_\mu{}^{A'}$ in~\eqref{eq:stueckelberg-fix-no-long-metric}, where $A'=2,\ldots,d-1$\,.
Together with their inverses $E^\mu{}_{A'}$ and the longitudinal vielbeine, they satisfy the following orthogonality and completeness relations \cite{Andringa:2012uz},
\begin{subequations}
\label{eq:long-trans-vielbeine-orth-compl}
    \begin{align}
        \tau_\mu{}^A \, \tau^\mu{}_B & = \delta^A_B\,,
            &%
        \tau_\mu{}^A \, E^\mu{}_{B'} =
        E_\mu{}^{A'} \tau^\mu{}_B & = 0\,,
        \\[2pt]
        E_\mu{}^{A'} \, E^\mu{}_{B'} & = \delta^{A'}_{B'}\,,
            &%
        \tau_\mu{}^A \, \tau^\nu{}_A + E_\mu{}^{A'} \, E^\nu{}_{A'} & = \delta_\mu^\nu\,.
    \end{align}
\end{subequations}
The resulting geometry is referred to as \emph{torsional string Newton--Cartan} (TSNC) geometry \cite{Bidussi:2021ujm,Harmark:2019upf, Bergshoeff:2021bmc, Gallegos:2020egk}.
In contrast to the usual Lorentzian geometry of general relativity, nonzero `intrinsic' torsion related to $d\tau^A$
arises naturally in these geometries for connections that are compatible with the NR geometric data.
Additionally, TSNC geometry has a codimension-two foliation structure, with leaves being the transverse sector.
See Fig.~\ref{fig:bread} for an illustration of such a foliation structure.
For this to be the case, the Frobenius integrability condition needs to hold, which in terms of the target-space one-forms $\tau^A = \tau_\mu{}^A \, dx^\mu$ is
\begin{equation}
  \label{eq:foliation-condition-tsnc}
  d\tau^A = \alpha^A{}_B \wedge \tau^B.
\end{equation}
This generalizes `regular' Newton--Cartan geometry with a single clock one-form $\check\tau=\check\tau_\mu dx^\mu$,
which corresponds to a foliation with $(d-1)$-dimensional spatial slices if the twistless torsional Newton--Cartan (TTNC) condition $\check\tau\wedge d\check\tau=0$ holds.
In the present `stringy' case, the condition~\eqref{eq:foliation-condition-tsnc} is equivalent to~\cite{Hartong:2021ekg}
\begin{equation}
  \label{eq:icp2}
  E^{\mu}{}_{A'} \, E^\nu{}_{B'} \, \p_{[\mu} \tau_{\nu]}{}^A = 0\,.
\end{equation}
As we will see later on, such foliation conditions play a role in the quantum consistency of the worldsheet theory.\footnote{%
    String foliation constraints also arise from the $1/c^2$ expansion of the relativistic string action~\cite{Hartong:2021ekg}, which we do not consider in this review.
}
More generally, conditions on $d\tau^A$
(which is related to torsion)
are sometimes also referred to as torsion conditions in the literature.
In particular, introducing a longitudinal spin connection $\Omega_\mu{}^A{}_B$, the condition
\begin{equation}
    D_{[\mu} \tau_{\nu]}{}^A
    = \pd_{[\mu} \tau_{\nu]}{}^A - \Omega_{[\mu}{}^A{}_B \, \tau_{\nu]}{}^B
    = 0
\end{equation}
has been proposed \cite{Andringa:2012uz}, which implies in particular that the foliation condition~\eqref{eq:foliation-condition-tsnc} holds.

After fixing the St\"{u}ckelberg symmetry as in~\eqref{eq:stueckelberg-fix-no-long-metric}, the sigma model action~\eqref{eq:cbsa} that describes NR strings becomes
\cite{Harmark:2019upf}
\begin{align} \label{eq:cbsaEM}
\begin{split}
    S & = \frac{1}{4\pi\alpha'} \int_\Sigma d^2 \sigma \, \Bigl( \sqrt{h} \, h^{\alpha\beta} \, \p_\alpha X^{\mu} \, \p_\beta X^{\nu} \, E_{\mu\nu} - i \, \epsilon^{\alpha\beta} \, \p_\alpha X^\mu \, \p_\beta X^\nu \, M_{\mu\nu} \Bigr) \\[2pt]
    & \,\quad + \frac{1}{4\pi\alpha'} \int_\Sigma d^2 \sigma \sqrt{h} \, \Bigl(\lambda \, \bar{\mathcal{D}} X^\mu \, \tau_\mu + \bar{\lambda} \, \mathcal{D} X^\mu \, \bar{\tau}_\mu + \alpha' \, \text{R} \, \Phi \Bigr)\,.
\end{split}
\end{align}
In the flat limit with $\tau_\mu{}^A = \delta_\mu^A$\,, $E_\mu{}^{A'} = \delta_\mu^{A'}$, and $M_{\mu\nu}= \Phi = 0$\,, this action reduces to the free action \eqref{eq:Sfree}.
The worldsheet global symmetries that act on $X^\mu$ are interpreted as local gauge symmetries of the target space.

Similar to how Lorentzian geometry can be seen as the gauging of the Poincar\'{e} algebra, we can use the resulting gauge symmetries to define the TSNC target space geometry.
The vielbein fields $\tau_\mu{}^A$ and $E_\mu{}^{A'}$ can be seen as gauge fields associated with the longitudinal translations $H_A$ and transverse translations $P_{A'}$.
The longitudinal Lorentz boost $J_{AB}=\epsilon_{AB} J$ and the transverse rotations $J_{A'B'}$ act on $\tau_\mu{}^A$ and $E_\mu{}^{A'}$ in the standard way.
In particular, the string Galilei boost, with generators $G_{AB'}$ and Lie group parameters $\Lambda^{A'}{}_A$\,, acts as
\begin{equation}
  \label{eq:tau-E-M-Gal-boosts}
  \delta_{\text{G}} \tau_\mu{}^A = 0\,,
  \qquad%
  \delta_{\text{G}} E_\mu{}^{A'} = \Lambda^{A'}{}_{A} \, \tau_\mu{}^A\,,
  \qquad%
  \delta_{\text{G}} M_{\mu\nu} = 2 \, \Lambda^{A'}{}_A \, \epsilon^{A}{}_B \, \tau_{[\mu}{}^B \, E_{\nu]}{}^{A'}.
\end{equation}
In addition, the string Galilei boost symmetry acts nontrivially on $\lambda$ and $\bar{\lambda}$\,.
Together, these symmetries form the \emph{string Galilei algebra}, whose commutators are given by \cite{Brugues:2004an}
\begin{subequations}
\label{eq:string-galilei-algebra}
  \begin{align}
    [J_{A'B'}, J_{C'D'}]
    &= \delta_{A'C'} \, J_{B'D'}
    - \delta_{B'C'} \, J_{A'D'}
    + \delta_{B'D'} \, J_{A'C'}
    - \delta_{A'D'} \, J_{B'C'}\,, \\[2pt]
    [J_{A'B'},P_{C'}]
    &= \delta_{A'C'} \, P_{B'} - \delta_{B'C'} \, P_{A'}\,,
        \hspace{1.5cm}%
    [J, H_A]
    = \epsilon^B{}_A \, H_B\,,
    \\[2pt]
    [J_{A'B'}, G_{CD'}]
    & = \delta_{A'D'} \, G_{CB'} - \delta_{B'D'} \, G_{CA'}\,,
        \qquad%
    [J,G_{AB'}]
    = \epsilon^C{}_A \, G_{CB'}\,,
    \\[2pt]
    [G_{AB'}, H_C]
    & = \eta_{AC} \, P_{B'}\,.
  \end{align}
\end{subequations}
The target-space fields also transform under diffeomorphisms as usual, and the antisymmetric field $M_{\mu\nu}$ transforms under a one-form gauge symmetry,
\begin{equation}
\label{eq:one-form-gauge-tr-M}
    \delta_\varepsilon M_{\mu\nu} = \pd_{\mu}\varepsilon_{\nu} - \pd_\nu \varepsilon_{\mu}\,.
\end{equation}
Finally, the sigma model \eqref{eq:cbsaEM} is invariant under a dilatation symmetry~\cite{Bergshoeff:2018yvt,Bergshoeff:2021bmc} that rescales the longitudinal vielbeine $\tau_\mu{}^A$ and the one-form fields $\lambda$ and $\bar\lambda$, while simultaneously shifting the dilaton field $\Phi$\,.
This action therefore describes classical strings moving in a TSNC geometry, corresponding to a gauged string Galilei algebra, augmented with a dilatation symmetry associated to $\Phi$ and a one-form gauge transformation associated to $M_{\mu\nu}$.

In the absence of the dilaton field, the Nambu-Goto form of the action can be obtained by integrating out the worldsheet zweibein $e_\alpha{}^a$ in the path integral, which leads to \cite{Andringa:2012uz}
\begin{equation}
  \label{eq:nrngf}
  S_\text{NG} = \frac{1}{4\pi\alpha'} \int_\Sigma d^2 \sigma \left( \sqrt{- \det \tau_{\alpha\beta}} \, \tau^{\alpha\beta} \, \p_\alpha X^\mu \, \p_\beta X^\nu \, E_{\mu\nu}
  - i \, \epsilon^{\alpha\beta} \, \p_\alpha X^\mu \, \p_\beta X^\nu \, M_{\mu\nu} \right).
\end{equation}
Here, $\tau_{\alpha\beta} = \p_\alpha X^\mu \, \p_\beta X^\nu \tau_{\mu\nu}$ is the pullback of $\tau_{\mu\nu} = \tau_\mu{}^A \, \tau_\nu{}^B \, \eta_{AB}$\,, corresponding to the induced metric on the worldsheet, and $\tau^{\alpha\beta}$ is its inverse.

An alternative formulation of the Polyakov form of the action~\eqref{eq:cbsaEM} can be obtained by
splitting off the longitudinal components in $M_{\mu\nu}$ in terms of an additional gauge field $m_\mu{}^A$,
\begin{equation} \label{eq:MmB}
  M_{\mu\nu} = B_{\mu\nu} + 2 \, m_{[\mu}{}^A \tau_{\nu]}{}^B \, \epsilon_{AB},
  \qquad
  \delta_\text{G} m_\mu{}^A = - \Lambda_{A'}{}^{A} \, E_\mu{}^{A'}.
\end{equation}
In doing so, we have absorbed the boost transformations in $m_\mu{}^A$, so that the remaining antisymmetric tensor $B_{\mu\nu}$ is invariant under boosts.
This results in the action \cite{Bergshoeff:2018yvt}
\begin{align} \label{eq:cbsaHB}
\begin{split}
    S & = \frac{1}{4\pi\alpha'} \int_\Sigma d^2 \sigma \, \Bigl( \sqrt{h} \, h^{\alpha\beta} \, \p_\alpha X^{\mu} \, \p_\beta X^{\nu} \, H_{\mu\nu} - i \, \epsilon^{\alpha\beta} \, \p_\alpha X^\mu \, \p_\beta X^\nu \, B_{\mu\nu} \Bigr) \\[2pt]
    & \quad\, + \frac{1}{4\pi\alpha'} \int_\Sigma d^2 \sigma \sqrt{h} \, \Bigl( \lambda \, \bar{\mathcal{D}} X^\mu \, \tau_\mu + \bar{\lambda} \, \mathcal{D} X^\mu \, \bar{\tau}_\mu + \alpha' \, \text{R} \, \Phi \Bigr)\,.
\end{split}
\end{align}
Here, we introduced the combination $H_{\mu\nu} = E_{\mu\nu} + 2 \, m_{(\mu}{}^A \, \tau_{\nu)}{}^B \, \eta^{\phantom{)}}_{AB}$, which is invariant under string Galilei boosts.
Since $B_{\mu\nu}$ no longer transforms under boosts, it is similar to the `standard' Kalb--Ramond field of relativistic strings, which transforms only under the $U(1)$ gauge symmetry $\delta_\xi B_{\mu\nu} = 2 \, \p_{[\mu} \xi_{\nu]}$\, and spacetime diffeomorphisms.
As such, this alternative parametrization separates the $B$-field on the one hand from the geometric data $H_{\mu\nu}\,$, $\tau_\mu{}^A$\, and $m_\mu{}^A$ on the other hand.
These two groups of variables then do not transform into each other under the string Galilei symmetries, akin to the case in relativistic string theory. However,
without imposing $\tau^\mu{}_A \, \tau^\nu{}_B \, B_{\mu\nu}=0$, this description of the target space variables reintroduces a St\"{u}ckelberg symmetry similar to~\eqref{eq:Stueckelberg}.
As a result, $H_{\mu\nu}$ and $B_{\mu\nu}$ can never be completely separated in any physical observable.
Still, the requirement of St\"{u}ckelberg symmetry can provide a useful check on computations in terms of this alternative parametrization.

\subsection{Torsional string Newton--Cartan geometry from a limit} \label{sec:tsncgfal}
In \S\ref{sec:zeroRegge}, we reviewed how NR string theory in flat spacetime arises as a zero Regge slope limit of relativistic string theory.
This limiting procedure can be directly generalized to strings propagating in arbitrary background fields.
For this, we start from the sigma model for relativistic string theory,
\begin{align} \label{eq:Shat}
\begin{split}
    \hat{S} = \frac{1}{4\pi\hat\alpha'} \int_\Sigma d^2 \sigma \, \sqrt{h} \, \left[ \mathcal{D} X^\mu \, \bar{\mathcal{D}} X^\nu \, \bigl( \hat{G}_{\mu\nu} + \hat{B}_{\mu\nu} \bigr) + \hat\alpha' \, R \, \hat{\Phi} \right].
\end{split}
\end{align}
Next, we introduce a set of longitudinal vielbeine $\tau_\mu{}^A$.
We then parametrize the relativistic Lorentzian background metric $\hat{G}_{\mu\nu}$, the Kalb--Ramond field $\hat{B}_{\mu\nu}$ and the dilaton $\hat{\Phi}$ using \cite{Andringa:2012uz}
\begin{equation}
  \label{eq:gbexp}
  \hat{G}_{\mu\nu} = c^2 \, \tau_{\mu\nu} + E_{\mu\nu}\,,
  \qquad%
  \hat{B}_{\mu\nu} = - c^2 \, \tau_\mu{}^A \tau_\nu{}^B \epsilon_{AB} + M_{\mu\nu}\,,
  \qquad%
  \hat{\Phi} = \Phi - \ln |c|\,,
\end{equation}
where $E_{\mu\nu} = \delta_{A'B'} E_\mu{}^{A'} E_\nu{}^{B'}$\,.
In the flat limit, with $\tau_\mu{}^A\to\delta_\mu^A$\,, $E_\mu{}^{A'} \rightarrow \delta_\mu^{A'}$ and $M_{\mu\nu} \rightarrow 0$\,, the relativistic background fields in \eqref{eq:gbexp} reduce to the choice of background fields~\eqref{eq:ncoslimit} in flat spacetime.
We take $\tau_\mu{}^A$, $E_\mu{}^A$ and $M_{\mu\nu}$ to be independent of the parameter $c$.
To be able to take the $c\to\infty$ limit on the worldsheet, we then introduce a pair of one-form fields $\lambda$ and~$\bar\lambda$, which allows us to rewrite the action~\eqref{eq:Shat} as
\begin{align} \label{eq:cbsa-repeat}
\begin{split}
    \hat{S} = \frac{1}{4\pi\alpha'} \int_\Sigma d^2 \sigma \, \sqrt{h} \, \Bigl\{ &  \mD X^{\mu} \, \bar\mD X^{\nu} \left( E_{\mu\nu} + M_{\mu\nu} \right) \\[2pt]
    &\quad + \lambda \, \bar{\mathcal{D}} X^\mu \, \tau_\mu + \bar{\lambda} \, \mathcal{D} X^\mu \, \bar{\tau}_\mu + \lambda \bar{\lambda} \, U + \alpha' \, \text{R} (h) \, \Phi \Bigr\}\,,
\end{split}
\end{align}
where we have identified $U=1/c^2$.
We then promote $U$ to be a functional coupling depending on $X^\mu$.
This corresponds precisely to the action~\eqref{eq:cbsa} with general marginal couplings that we introduced at the beginning of this section.
As a result, we see that sending $c\to\infty$ in the relativistic theory~\eqref{eq:Shat} using the parametrization~\eqref{eq:gbexp} of the background fields removes the $\lambda\bar\lambda \, U$ term in the worldsheet action.
This produces the sigma model~\eqref{eq:cbsaEM} for NR string theory in arbitrary backgrounds.
Up to rescalings, the $c\to\infty$ (or $U\to0$) limit is equivalent to the zero Regge slope limit that we considered in the previous section for flat spacetime,
where the parametrizations in~\eqref{eq:gbexp} reduce to the ones in~\eqref{eq:ncoslimit} with a critical $B$-field.

While we are able to make sense of the $c\to\infty$ limit on the worldsheet,
this limit seems singular
from the perspective of the relativistic NS-NS geometric data $\hat{G}_{\mu\nu}$, $\hat{B}_{\mu\nu}$ and~$\hat{\Phi}$ in~\eqref{eq:gbexp}.
This is not surprising, since the NR string sigma model~\eqref{eq:cbsaEM} that results from the limit does not couple to a relativistic NS-NS geometry but to TSNC geometry, as we discussed above.
It is also important to understand the NR limit directly on the level of the spacetime geometry.
We will now review two different methods that both show how one can obtain TSNC geometry from Lorentzian geometry, without relying on the worldsheet theory.
The first method starts from the description of Lorentzian geometry in terms of a gauging of the Poincar\'{e} algebra,
which can be extended to include the Kalb--Ramond field.
The second method uses double field theory, which incorporates both the metric and the Kalb--Ramond field into a single $O(d,d)$-covariant metric $\mH_{AB}$\,.
Both of these setups can be used to consistently describe the $c\to\infty$ limit of the target space geometry. For simplicity, we will not consider the dilaton in the following discussion.

\subsubsection{Algebra gauging}
\label{sssec:algebra-gauging}
From an algebraic perspective, Lorentzian geometry can be obtained from a gauging of the Poincar\'{e} algebra.
This gauging associates the Lorentzian vielbeine $E_\mu{}^\hA$, where we have $\hA= 0,1,\ldots d-1$\,, to the translation generators $P_\hA$\,, while the spin connection $\Omega_\mu{}^{\hA\hB}$ is associated to the Lorentz boost generators $M_{\hA\hB}$\,.
One can incorporate the Kalb--Ramond field into this construction by adding an additional set of generators $Q_\hA$ to the Poincar\'{e} algebra, which satisfy the same commutation relation as the translation generators \cite{Harmark:2019upf,Bidussi:2021ujm}.
The associated fields, denoted by $\Pi_\mu{}^\hA$, can therefore be thought of as an `additional' set of vielbeine.
With this, the total connection is
\begin{equation}
    \mathcal{A}_\mu
    = E_\mu{}^\hA \, P_\hA + \frac{1}{2} \, \Omega_\mu{}^{\hA\hB} \, M_{\hA\hB} + \Pi_\mu{}^\hA \, Q_\hA\,.
\end{equation}
The Lorentzian tangent space metric $\hat{G}_{\mu\nu}$ is constructed from the vielbeine $E_\mu{}^\hA$ using the Minkowski frame metric $\eta_{\hA\hB}$.
Likewise, we use the additional vielbeine $\Pi_\mu{}^\hA$ to parametrize the relativistic Kalb--Ramond field,
\begin{equation}
  \label{eq:bfield-vielbein-param-rel}
  \hat{G}_{\mu\nu} = E_{\mu}{}^\hA \, E_{\nu}{}^\hB \, \eta_{\hA\hB}\,,
    \qquad%
  \hat{B}_{\mu\nu} = E_{[\mu}{}^\hA \, \Pi_{\nu]}{}^\hB \, \eta_{\hA\hB}\,.
\end{equation}
While the $\Pi_\mu{}^\hA$ fields initially have $d^2$ degrees of freedom, this parametrization of $\hat{B}_{\mu\nu}$ is invariant under shifting $\Pi_\mu{}^\hA\to P^\hA{}_\hB \, E_\mu{}^\hB$ for any symmetric $P_{\hA\hB}$\,, which leaves the correct amount of degrees of freedom for an antisymmetric tensor.
The gauge transformations associated to the Lorentz boosts correspond to local Lorentz transformations, while the translations can be related to diffeomorphisms.
Similarly, the transformations associated to $Q_\hA$ correspond to one-form gauge transformations.
Using the appropriate identifications, this can be done without any constraints on the torsion of the geometry.

To implement the NR limit, we split the frame indices $\hat{A}=(A,A')$ into longitudinal and transverse indices and introduce the reparametrization \cite{Bidussi:2021ujm}
\begin{equation} \label{eq:EPi}
  E_\mu{}^A = c \left(\tau^A_\mu + \frac{1}{2 \, c^2} \, \epsilon^A{}_B \, \pi_\mu{}^B \right),
    \qquad%
  \Pi_\mu{}^A = - c \, \epsilon^A{}_B \left( \tau_\mu{}^B - \frac{1}{2\,c^2} \, \epsilon^B{}_C \, \pi_\mu{}^C \right).
\end{equation}
The transverse vielbeine correspond to $E_\mu{}^{A'}$ and $\pi_\mu{}^{A'} = \Pi_\mu{}^{A'}$.
Using the parametrizations in~\eqref{eq:bfield-vielbein-param-rel}, this results in the following expansions,~\footnote{A related expansion also appears in \cite{Bergshoeff:2019pij}.}
\begin{subequations}
\label{eq:metric-bfield-vielbein-expansion}
  \begin{align}
    \hat{G}_{\mu\nu}
    &= c^2 \, \tau_{\mu\nu}
    + \tau_{(\mu}{}^A \, \pi_{\nu)}{}^B \, \epsilon^{}_{AB}
    + E_{\mu\nu}
    + \OO(1/c^2)\,,
    \\[2pt]
    \hat{B}_{\mu\nu}
    &= - c^2 \, \tau_{\mu}{}^A \, \tau_{\nu}{}^B \, \epsilon^{}_{AB} - \tau_{[\mu}{}^A \, \pi_{\nu]}{}^B \, \eta_{AB} + M_{\mu\nu}
    + \OO(1/c^2)\,.
  \end{align}
\end{subequations}
Here, we recover the transverse metric $E_{\mu\nu} = \delta_{A'B'} \, E_\mu{}^{A'} \, E_\nu{}^{B'}$ and the antisymmetric tensor~$M_{\mu\nu}$ that were introduced in
\eqref{eq:stueckelberg-fix-no-long-metric}.
The latter is now parametrized as
\begin{equation} \label{eq:mEpi}
  M_{\mu\nu}
  = \tau_{[\mu}{}^A \, \pi_{\nu]}{}^B \, \eta^{}_{AB}
  + E_{[\mu}{}^{A'} \, \pi_{\nu]}{}^{B'} \, \delta^{}_{A'B'}\,.
\end{equation}
In the $c\to\infty$ limit, the redefinition~\eqref{eq:EPi} corresponds to an $\dot{\text{I}}$n\"{o}n\"{u}--Wigner-type contraction of the relativistic algebra.
We can then gauge the resulting \emph{F-string Galilei} algebra~\cite{Bidussi:2021ujm} to obtain the local transformations associated to the geometry.
These local transformations, which include both the string Galilei boosts~\eqref{eq:tau-E-M-Gal-boosts} and the one-form gauge transformations~\eqref{eq:one-form-gauge-tr-M}, can be derived from the F-string Galilei algebra without any restrictions on the torsion of the geometry.
The result is the spacetime TSNC geometry that the nonrelativistic string sigma model~\eqref{eq:nrngf} couples to.

\subsubsection{Double field theory}
\label{sssec:dft}
An alternative approach to parametrizing non-Lorentzian geometries comes from double field theory (DFT) \cite{Lee:2013hma,Ko:2015rha,Morand:2017fnv}.
This formalism was originally intended to provide a manifestly T-duality covariant description of the geometry that relativistic strings couple to (see for example~\cite{Aldazabal:2013sca} for a review).
In relativistic string theory, the DFT formalism unifies the Lorentzian metric and the Kalb--Ramond field into a single \emph{generalized} metric,
\begin{equation}
\label{eq:dft-gen-metric-riem-param}
    \mH_{MN} = \begin{pmatrix}
        \hat{G}^{\mu\nu}
        &\,\,\,\, - \hat{G}^{\mu\rho} \, \hat{B}_{\rho\nu}
        \\[4pt]
        \hat{B}_{\mu\rho} \, \hat{G}^{\rho\nu}
        &\,\,\,\, \hat{G}_{\mu\nu} - \hat{B}_{\mu\rho} \, \hat{G}^{\rho\sigma} \, \hat{B}_{\sigma\nu}
    \end{pmatrix}.
\end{equation}
The indices $M,N$ are $2d$-dimensional and they are raised and lowered using the $O(d,d)$ metric
\begin{equation}
    \mJ_{MN}
    = \begin{pmatrix}
        0 &\,\, \delta^\mu{}_\nu
        \\
        \delta_\mu{}^\nu &\,\, 0
    \end{pmatrix}.
\end{equation}
The doubled coordinates $X^M = (\tilde{X}_\mu, X^\mu)$ incorporate both
the conventional coordinates
$X^\mu$ and the `dual' coordinates $\tilde{X}_\mu$\,.
For consistency of the theory, one needs to impose a section condition, which is commonly solved by requiring the fields to be independent of the dual coordinates $\tilde{X}_\mu$.
Together with the covariant dilaton $e^{-2\hat{\Phi}}\,|\hat{G}|^{1/2}$\,, one can then construct $O(d,d)$-covariant `doubled' actions for both the string sigma model and the target-space effective action (see for example~\cite{Lee:2013hma,Angus:2018mep}).
The generalized metric given in~\eqref{eq:dft-gen-metric-riem-param} is symmetric, and its inverse is obtained simply by raising its indices with~$\mJ_{MN}$.
We can encode these properties in an $O(d,d)$ covariant way using
\begin{equation}
    \label{eq:dft-gen-metric-def-relations}
    \mH_{MN} = \mH_{NM},
    \qquad
    \mH_M{}^P \, \mH_N{}^Q \, \mJ_{PQ} = \mJ_{MN}.
\end{equation}
Remarkably, the particular combinations of the Lorentzian metric and Kalb--Ramond field that enter in the parametrization~\eqref{eq:dft-gen-metric-riem-param} combine in such a way that its $c\to\infty$ limit using the expansion~\eqref{eq:metric-bfield-vielbein-expansion} is nonsingular.
In this limit, we obtain~\cite{Morand:2017fnv,Blair:2020gng}
\begin{equation}
\label{eq:dft-gen-metric-tsnc-param}
    \mH_{MN}
    = \begin{pmatrix}
        E^{\mu\nu}
        &\quad - E^{\mu\rho} \, M_{\rho\nu} + \tau^{\mu}{}_{A} \, \tau_\nu{}^B \, \epsilon^A{}_{B}\,,
        \\[4pt]
        M_{\mu\rho} \, E^{\rho\nu} - \tau_\mu{}^A \, \tau^\nu{}_B \, \epsilon_A{}^B
        &\quad E_{\mu\nu} - M_{\mu\rho} E^{\rho\sigma} M_{\sigma \nu}
        - 2 \, \tau_{(\mu}{}^A \, M_{\nu)\rho} \, \tau^\rho{}_B \, \epsilon_A{}^B
    \end{pmatrix}.
\end{equation}
Clearly, this generalized metric no longer corresponds to a Lorentzian metric, since its top left block $E^{\mu\nu}$ is now degenerate.
However, if one considers the doubled description of the target space geometry
as fundamental, one can take~\eqref{eq:dft-gen-metric-def-relations} as the \emph{defining} relations of the geometry.
From this perspective, the parametrization~\eqref{eq:dft-gen-metric-tsnc-param} in terms of TSNC variables therefore constitutes a valid generalized metric in the doubled actions, since these defining relations still hold, even though the generalized metric can no longer be related to relativistic NS-NS geometry.

The most general solution of the defining equations~\eqref{eq:dft-gen-metric-def-relations} for $\mH_{AB}$ leads to a top left $d\times d$ block with $(n,\bar{n})$ chiral and antichiral lightlike vectors, and a string sigma model can be constructed for each case~\cite{Morand:2017fnv}.
The TSNC geometry discussed above corresponds to the case $(n,\bar{n})=(1,1)$.
Having $n=\bar{n}$ appears to be necessary for zero central charge in the BRST algebra on the string worldsheet~\cite{Park:2020ixf}.
The framework of double field theory has also been used to study target space effective actions~\cite{Cho:2019ofr,Gallegos:2020egk}, the relation between the $\lambda\bar\lambda$ and $T\bar{T}$ deformations~\cite{Blair:2020ops}, worldsheet symmetry algebras~\cite{Blair:2020gng} and supersymmetric string actions~\cite{Park:2016sbw,Blair:2019qwi}.
Similar limits and NR parametrizations have also been considered in exceptional field theory, which generalizes the manifest duality-invariance of string theory to M-theory~\cite{Berman:2019izh,Blair:2021waq}.

\subsection{Torsional string Newton--Cartan geometry from null reduction}
\label{ssec:null-red}

NR string actions can also be obtained from a null reduction in a relativistic background with a lightlike isometry~\cite{Harmark:2017rpg}.
Starting from a $d$-dimensional Lorentzian manifold with a lightlike isometry, we choose adapted coordinates $x^\mu = (u,x^\hmu)$ such that the lightlike isometry is generated by~$\pd_u$.
Then we can write the associated metric as
\begin{equation}
    \label{eq:metric-null-isometry-tnc}
    ds^2
    = 2 \, \tau_\hmu \, dx^\hmu \left( du - m_\hnu \, dx^\hnu \right) + E_{\hmu\hnu} \, dx^\hmu dx^\hnu.
\end{equation}
Together with $u$, the $d-1$ coordinates $x^{\hmu}$ form a chart of the target-space manifold.
Additionally, we decompose the Kalb--Ramond field in the components $b_\hmu=B_{u\hmu}$ and $B_{\hmu\hnu}$.
In such a background, the relativistic Polyakov action is given by
\begin{equation}
\begin{split}
    \label{eq:Shat-null}
    \hat{S}
    &= \frac{1}{4\pi\alpha'} \int d^2\sigma
    \left[
        \sqrt{h} \, h^{\alpha\beta} \left(E_{\alpha\beta} - 2\tau_{(\alpha} m_{\beta)}\right)
        - i\, \epsilon^{\alpha\beta} B_{\alpha\beta}
    \right]
    \\
    &{}\qquad\qquad
    + \frac{1}{2\pi\alpha'} \int d^2\sigma
        \left( \sqrt{h} \, h^{\alpha\beta} \, \tau_\beta
        - i\,\epsilon^{\alpha\beta} \, b_\beta \right) \pd_\alpha X^u.
\end{split}
\end{equation}
As we have discussed in \S\ref{sec:smbf}, $X^u$ and $X^\hmu$ are worldsheet functions that parametrize the embedding of the worldsheet in terms of the target space coordinates $u$ and $x^\hmu$.
Pullbacks such as $\tau_\alpha=\pd_\alpha X^\hmu \tau_\hmu$ are constructed using the embedding coordinates $X^\hmu$.
To implement a null reduction, we also require that the momentum of the string in the $u$-direction is conserved off shell.
For this, we introduce a Lagrange multiplier~$A_\alpha$ that imposes a relation between the momentum current $P_u^\alpha = \pd \hat{\mathcal{L}} / \pd( \pd_\alpha X^u)$ and a closed one-form $d\eta$ on the worldsheet~\cite{Harmark:2017rpg,Harmark:2018cdl,Harmark:2019upf},
\begin{equation}
\begin{split}
    \label{eq:S-tnc-string}
    S'
    &= \frac{1}{4\pi \alpha'} \int d^2\sigma
    \left[
        \sqrt{h} h^{\alpha\beta} \left(E_{\alpha\beta} - 2\tau_{(\alpha} m_{\beta)}\right)
        -i \epsilon^{\alpha\beta} B_{\alpha\beta}
    \right]
    \\
    &{}\qquad\qquad
    + \frac{1}{2\pi\alpha'} \int d^2\sigma
        \left[ \sqrt{h} \, h^{\alpha\beta} \tau_\beta
        -i \epsilon^{\alpha\beta} \left(b_\beta + \pd_\beta \eta\right) \right] A_\alpha.
\end{split}
\end{equation}
In this action, the equations of motion for $\eta$ imply that the one-form $A_\alpha \, d\sigma^\alpha$ is closed.
As a result, we can recover the original lightlike direction through $A_\alpha = \pd_\alpha X^u$ and we see that our action is equivalent to the relativistic string action~\eqref{eq:Shat-null}.
Alternatively, we can interpret~$\eta$ itself as an embedding coordinate that is \emph{dual} to $X^u$.
We work in a sector of fixed total momentum in the $u$-direction, and interpret $\eta$ as a compact target-space direction along which the string has a fixed winding mode.
Then the constraint imposed by the Lagrange multipliers $A_\alpha$ imply that the total momentum of the string in the $u$-direction is mapped to the string winding in $\eta$\,.
For this reason, the dual NR string winds exactly once in~$\eta$\,, and the periodicity of this direction is determined by the original lightlike momentum in $X^u$.

The action~\eqref{eq:S-tnc-string} describes strings coupled to a torsional Newton--Cartan geometry, described by the fields $\tau_\hmu$\,, $E_{\hmu\hnu}$, together with the background fields $m_\hmu$, $B_{\hmu\hnu}$ and $b_\hmu$ as well as the $\eta$ direction along which the string winds.
It is equivalent to the worldsheet action~\eqref{eq:cbsaEM} introduced above after the following identifications~\cite{Harmark:2018cdl,Harmark:2019upf,Bidussi:2021ujm}
\begin{equation}
\label{eq:tnc-snc-identification}
\begin{gathered}
    x^\mu = (x^\hmu,v)\,,
    \qquad
    \tau_\mu{}^0 = (\tau_\hmu,0)\,,
    \qquad
    \tau_\mu{}^1 = (b_\hmu,1)\,,
    \\
    M_{\hmu\hnu} = B_{\hmu\hnu} + 2 m_{[\hmu} \, b_{\hnu]}\,,
    \qquad
    M_{\hmu v} = m_\hmu\,.
\end{gathered}
\end{equation}
The last line defines the antisymmetric tensor $M_{\mu\nu}$, where $x^\hmu$ and $v$ form a chart of the dual target space manifold.
Similarly, the symmetric tensor $E_{\hmu\hnu}$ is extended to $E_{\mu\nu}$ by setting $E_{vv} = E_{\hmu v}=0$.
The Lagrange multipliers $A_\alpha$ are related to the one-forms $\lambda$ and $\bar\lambda$ in the action~\eqref{eq:cbsaEM}, and the local boost symmetries and one-form gauge transformations of the background can also be recovered.
As such, null reduction provides an alternative perspective on the NR string action and the TSNC geometry that we previously obtained from a $c\to\infty$ limit.
Likewise, in addition to the limiting procedure we discussed in \S\ref{sssec:dft} above, non-Riemannian parametrizations of generalized metrics can be obtained from a generalized metric in the relativistic parametrization~\eqref{eq:dft-gen-metric-riem-param} using an $O(d,d)$ transformation along a lightlike isometry~\cite{Berman:2019izh,Blair:2019qwi}.

In the above, we have considered a single lightlike momentum sector of strings in the relativistic background~\eqref{eq:metric-null-isometry-tnc}.
The lightlike isometry in the relativistic background corresponds to an isometry of the spatial longitudinal direction $v$ in the dual TSNC geometry, and the momentum mode of the string along $u$ is translated to a single winding mode of the nonrelativistic string along $v$.
For related constructions of nonrelativistic particle or field theory actions through null reduction, the lightlike direction $u$ is typically taken to be noncompact.
On the other hand, the T-duality relation between NR strings and the DLCQ of relativistic strings that was discussed in \S\ref{ssec:flat-nr-closed-strings} would require that $u$ is a compact lightlike direction.

\section{Effective Field Theories from Nonrelativistic Strings} \label{sec:eftns}

Next, we review the RG analysis of the sigma model~\eqref{eq:cbsaEM} where the classical value for the background field associated with the $\lambda\bar{\lambda}$ operator is tuned to zero.
This analysis will allow us to construct an effective Newton-like theory of gravity in the target space.
In general, the $\lambda\bar{\lambda}$ operator is generated by log-divergent loop corrections.
As a result, this operator will have to be included in the spectrum in order for the OPEs to be closed, and it would then deform the theory back to the full relativistic string theory.
To counteract this, extra global symmetries are imposed in the worldsheet sigma model such that this operator can be prevented from being generated quantum-mechanically.
Such worldsheet global symmetries correspond to additional spacetime gauge symmetries that restrict (part of) the torsion of the target-space TSNC geometry.

We discuss different proposals for symmetry algebras that have been used to construct renormalizable interacting worldsheet QFTs that describe NR strings in background fields.
Imposing Weyl invariance at the quantum level, the vanishing beta-functionals of background fields determine the spacetime equations of motion that govern the dynamics of the target-space TSNC geometry. This is analogous to how the (super)gravity equations of motion arise in relativistic string theory.
With such worldsheet symmetries, NR string theory can be studied in a self-contained way, without referring to the full relativistic string theory.
We also comment on recent progress on supersymmetrizations of NR string theory and its relation to the modified symmetry algebras.
Finally, we review recent progress on the RG calculation of the worldsheet theory for NR open strings.
This analysis of conformal anomalies gives the NR analog of Dirac-Born-Infeld (DBI) action that describes the low energy dynamics of D-branes.
The distinction between different NR limits considered in the literature that lead to extended $p$-brane objects in NR string/M-theory will also be discussed.

\subsection{Beta functions of the worldsheet theory}
\label{eq:rnrst}

After reviewing classical aspects of NR string sigma models and the associated target-space geometry, we now investigate the quantum behavior of the worldsheet QFT.
We first approach this by taking a limit of the beta functionals of relativistic string theory.
Next, we review different constructions of self-contained NR string theories in curved spacetime that are defined by renormalizable sigma models.

\subsubsection{Target-space gravity from a limit} \label{eq:tsgl}
The self-consistency of string theory requires the classical worldsheet Weyl invariance to be preserved at the quantum level. This sets all beta-functionals of the background fields in the string sigma model to zero. The vanishing beta-functionals determine the target-space equations of motion that govern the dynamics of the target-space geometry. In relativistic string theory, this procedure leads to the spacetime supergravity equations of motion.

We already learned that the bosonic string sigma model \eqref{eq:Shat} describes strings propagating in NR geometries only if $U=0$\,, corresponding to the limit $c = U^{-1/2} \rightarrow \infty$\,.
This limit has been applied to
the spacetime equations of motion determined by the vanishing beta-functionals \cite{Bergshoeff:2021bmc}. It is shown that the dynamics of the NR target space arises from the $c \rightarrow \infty$ limit of the NS-NS gravity in relativistic string theory. At the lowest order in $\alpha'$, this defines the so-called dilatation-invariant string Newton-Cartan gravity \cite{Bergshoeff:2021bmc}.
The associated target-space gravity action has been studied in \cite{Bergshoeff:2021bmc} and also from a DFT point of view in~\cite{Gallegos:2020egk}.
Since these results are rather involved, we will only show them in a simplified model where the torsion is set to zero, which is discussed in \S\ref{sec:nce}.

The supersymmetric generalization of the parametrization \eqref{eq:gbexp} has been studied in various works. See \cite{Gomis:2004pw, Gomis:2005bj,Gomis:2005pg, Bergshoeff:2021tfn} for examples, and we will follow closely the recent paper  \cite{Bergshoeff:2021tfn} for the state of the art. In addition to the reparametrization of the metric, Kalb-Ramond, and dilaton field in \eqref{eq:gbexp}, one also needs to reparametrize the fermionic fields, including the gravitino~$\hat{\Psi}_\mu$ and the dilatino $\hat{\lambda}$\,, such that
\begin{equation}
  \hat{\Psi}_+ = c^{1/2} \, \psi_+ + c^{-1/2} \, \psi_-\,,
  \qquad%
  \hat{\lambda} = c^{1/2} \, \lambda_+ + c^{-1/2} \, \lambda_-\,,
\end{equation}
where $\psi_\pm$ and $\lambda_\pm$ are worldsheet chirality projected spinors. The $c \rightarrow \infty$ limit is nonsingular if the following geometric constraints hold:
\begin{equation}
  \label{eq:hztcs}
  E^\mu{}_{\!A'} \, E^\nu{}_{\!B'} \, \p_{[\mu} \bar{\tau}_{\nu]} = 0\,,
  \qquad%
  E^\mu{}_{\!A'} \, \tau^\nu \, \p_{[\mu} \bar{\tau}_{\nu]} = 0\,.
\end{equation}
One can choose whether these constraints are imposed on $d\tau$ or $d\bar{\tau}$\,. The condition~\eqref{eq:hztcs} is required for the supersymmetric transformation rules of $\psi_\pm$ and $\lambda_\pm$ to remain finite. It is interesting to note that these constraints only impose half of the integrability conditions~\eqref{eq:foliation-condition-tsnc} required by the codimension-two foliation. We will see later in \S\ref{sec:nce} that the same torsion condition~\eqref{eq:hztcs} shows up in bosonic sigma models once appropriate worldsheet symmetries are imposed.
Such a $c \rightarrow \infty$ limit has been applied to ten-dimensional $\mathcal{N} = 1$ supergravity, which leads to a supersymmetric generalization of TSNC geometry with NR spacetime supersymmetry~\cite{Bergshoeff:2021tfn}.

\subsubsection{Quantum corrections and renormalizability}
It is also insightful to apply the $U = c^{-2} \rightarrow 0$ limit to the beta-functionals of the sigma model~\eqref{eq:Shat} that describes relativistic strings propagating in Lorentzian geometries, before committing to the conformal fixed point \cite{Yan:2021lbe}.
This leads to the RG structure of the sigma model action
\eqref{eq:cbsaEM}, evaluated around the physical value $U = 0$\,.
Moreover, the Weyl invariance of sigma models in torsional Newton-Cartan backgrounds that we discussed in \S\ref{ssec:null-red} has also been studied directly using the worldsheet QFT
in~\cite{Gallegos:2019icg}.
It is shown in \cite{Gomis:2019zyu, Gallegos:2019icg} that the $\lambda\bar{\lambda}$ operator receives log-divergent quantum corrections, and its functional coupling $U$ receives nontrivial RG flows just like all other background fields. At the lowest order in $\alpha'$, the beta-functional of $U$ gives
\begin{equation}
  \label{eq:betaU}
  \beta^{(U)} \Big|_{U = 0} = \alpha' \, E^{\mu\rho} \, E^{\nu\sigma} \, \p^{}_{[\mu} \tau^{}_{\nu]} \, \p^{}_{[\rho} \bar{\tau}^{}_{\sigma]} + O(\alpha'{}^2)\,.
\end{equation}
Here, we defined $E^{\mu\nu} = E^\mu{}_{A'} E^{\nu A'}$.
This beta-functional \eqref{eq:betaU} is found by evaluating the quantum loop corrections directly using the sigma model \eqref{eq:cbsaEM} with $U=0$ in \cite{Gomis:2019zyu, Gallegos:2019icg}, and also from considering the NR limit \cite{Yan:2021lbe}.
Therefore, the sigma model \eqref{eq:cbsaEM} is not renormalizable unless a $\lambda\bar{\lambda}$ counterterm is included.
This issue does not invalidate the $U \rightarrow 0$ limit that we take in string theory: because the worldsheet has to be conformal for string theory to be self-consistent, all the beta-functionals are required to vanish. It is therefore consistent to tune $U = 0$ at the conformal point, together with the condition
\begin{equation}
  \label{eq:betaU0}
  \beta^{(U)} \Big|_{U = 0} = 0\,.
\end{equation}
At the lowest order in $\alpha'$, \eqref{eq:betaU} and \eqref{eq:betaU0} lead to the geometric constraints
\begin{equation}
    E^{\mu\rho} \, E^{\nu\sigma} \, \p_{[\rho} \tau_{\sigma]} = 0\,,
        \quad \text{or} \quad%
    E^{\mu\rho} \, E^{\nu\sigma} \p_{[\rho} \bar{\tau}_{\sigma]} = 0\,.
\end{equation}
These constraints are the lightlike components of the foliation condition \eqref{eq:icp2} and coincide with the first condition in \eqref{eq:hztcs} obtained from requiring the supersymmetry transformations to be finite.
The resulting NR gravity is therefore a zero $U$ solution to the (super)gravity equations of motion in relativistic string theory.

There are two perspectives on how \eqref{eq:betaU0} should be treated \cite{Yan:2021lbe}. In the first perspective \cite{Harmark:2017rpg, Harmark:2018cdl, Gallegos:2019icg, Harmark:2019upf, Bergshoeff:2021bmc}, we solve \eqref{eq:betaU0} together with other target-space equations of motion perturbatively, order by order in $\alpha'$. This defines target-space non-Lorentzian gravity with higher derivative corrections, but with the potential problem that solutions to the target-space equations of motion at lower orders in $\alpha'$ might not be extendable to higher orders without introducing a nonzero $U$\,.
In the second perspective \cite{Bergshoeff:2018yvt, Bergshoeff:2019pij, Gomis:2019zyu, Yan:2019xsf}, we are interested in identifying the conditions under which quantum corrections to $\lambda\bar{\lambda}$ vanish at all loops, which means that
\eqref{eq:betaU0} holds nonperturbatively at all loops, such that NR string theory is defined by a renormalizable worldsheet QFT.
This is achieved by extending the string Galilei algebra using additional generators, whose realization on the worldsheet imposes additional geometric constraints in the target-space geometry. These geometric constraints protect $\lambda\bar{\lambda}$ from being generated by quantum corrections, and thus lead to a self-contained notion of NR string theory that is free from deformations towards relativistic string theory.

\subsubsection{Extensions of string Galilei symmetries} \label{sec:nce}
In the following, we review several proposals for extended worldsheet global symmetry algebras, whose realization on the sigma model \eqref{eq:cbsaHB} results in different constraints on the longitudinal vielbein field $\tau_\mu{}^A$ that restrict the torsion of the TSNC geometry.
Once such constraints are imposed, the sigma model \eqref{eq:cbsaEM} is protected against any $\lambda\bar{\lambda}$ deformation and becomes renormalizable.

In \S\ref{sec:smbf}, we showed that the sigma model \eqref{eq:cbsaEM}
is invariant under the symmetry transformations
that form the string Galilei algebra \eqref{eq:string-galilei-algebra}, which arises as a contraction of the Poincar\'{e} algebra.
Recall that the string Galilei algebra consists of generators associated with longitudinal and transverse translations that we refer to as $H_A$ and $P_{A'}$\,, respectively, as well as the string Galilei boosts $G_{AA'}$\,. Notably, $P_{A'}$ and $G_{AA'}$ commute in \eqref{eq:string-galilei-algebra}.

In analogy to how the Galilei algebra can be extended to the Bargmann algebra in the particle case, it is shown in \cite{Brugues:2004an, Brugues:2006yd, Andringa:2012uz} that the string Galilei algebra can be extended to the \emph{string Bargmann algebra} by introducing a generator $Z_A$ such that
\begin{equation}
  [P_{A'}\,, G_{B'A}] = \delta_{A'B'} \, Z_A\,.
\end{equation}
This new generator $Z_A$ is the stringy version of the central charge corresponding to the conserved particle number in the Bargmann algebra. In contrast, because $Z_A$ carries a longitudinal index and hence does not commute with the longitudinal Lorentz boost, $Z_A$ is noncentral in the string Bargmann algebra.
One way to close the algebra is by including an additional generator $Y$ that arises from the commutator of $G_{AA'}$ and $G_{BB'}$.
This second extension will not play any important role in the following discussions, and we refer the readers to \cite{Bergshoeff:2019pij,Harmark:2018cdl} for discussions on the $Y$ extension and its generalizations.\footnote{
    Another way of including a $Z_A$ generator in the string Galilei algebra is provided by the F-string Galilei algebra mentioned in Section~\ref{sssec:algebra-gauging}.
    This algebra also includes generators corresponding to the one-form gauge transformations, and because of that it evades the need for introducing the additional extension $Y$.
    However, its realization on the worldsheet does not impose any constraints on the torsion, and it is therefore less suitable for the purposes of the current discussion.
}

The target-space gauge transformations corresponding to the string Bargmann algebra are most naturally realized on the variables $\tau_\mu{}^A$, $E_\mu{}^{A'}$, $m_\mu{}^A$ and $B_{\mu\nu}$ in the action~\eqref{eq:cbsaHB}.
In terms of these variables, the boost transformations remain the same as in \eqref{eq:MmB}, and the $Z_A$ symmetry only acts nontrivially on $m_\mu{}^A$,
\begin{equation}
    \delta m_\mu{}^A = D_\mu \sigma^A = \p_\mu \sigma^A + \Omega_\mu{}^{AB} \, \sigma_B\,,
\end{equation}
where we have used $\sigma^A$ to denote the parameter for $Z_A$ gauge transformations.~\footnote{This should not be confused with the worldsheet coordinates $\sigma^\alpha$.}
Additionally, the derivative $D_\mu$ is covariant with respect to the spin connection $\Omega_\mu{}^{AB}$ associated with the longitudinal Lorentz boost.
Recall that without imposing the condition $\tau^\mu{}_A \, \tau^\nu{}_B \, B_{\mu\nu}=0$, the theory has the St\"{u}ckelberg symmetry~\eqref{eq:Stueckelberg}, but the invariance under this symmetry can provide useful consistency checks on results in the quantum theory.
One can obtain the target-space variables by gauging the string Bargmann algebra, supplemented with a general antisymmetric $B_{\mu\nu}$ field, so that the latter only transforms under the $U(1)$ gauge symmetry $\delta_\xi B_{\mu\nu} = 2 \, \p_{[\mu} \xi_{\nu]}$ and spacetime diffeomorphisms.
From this perspective, $m_\mu{}^A$ is the gauge field associated to the $Z_A$ generator.

Requiring that the NR string sigma model is invariant under the background field transformations generated by the string Bargmann symmetries reproduces the action \eqref{eq:cbsaHB},
which we repeat below for convenience:
\begin{align} \label{eq:actionhb}
\begin{split}
    S & = \frac{1}{4\pi\alpha'} \int_\Sigma d^2 \sigma \, \Bigl( \sqrt{h} \, h^{\alpha\beta} \, \p_\alpha X^{\mu} \, \p_\beta X^{\nu} \, H_{\mu\nu} - i \, \epsilon^{\alpha\beta} \, \p_\alpha X^\mu \, \p_\beta X^\nu \, B_{\mu\nu} \Bigr) \\[2pt]
    & \quad\, + \frac{1}{4\pi\alpha'} \int_\Sigma d^2 \sigma \sqrt{h} \, \Bigl( \lambda \, \bar{\mathcal{D}} X^\mu \, \tau_\mu + \bar{\lambda} \, \mathcal{D} X^\mu \, \bar{\tau}_\mu + \alpha' \, \text{R} \, \Phi \Bigr).
\end{split}
\end{align}
Here, $H_{\mu\nu} = E_{\mu\nu} + 2 \, m_{[\mu}{}^A \, \tau_{\nu]}{}^B \, \eta_{AB}$ is invariant under string Galilei boosts but not under the $Z_A$ transformations.
For the action~\eqref{eq:actionhb} to be invariant under the $Z_A$ transformations, we must impose the zero-torsion condition on $\tau_\mu{}^A$
\cite{Andringa:2012uz},~\footnote{See \cite{Bergshoeff:2018vfn} for an example where the zero-torsion condition \eqref{eq:ztc} arises as an equation of motion from a spacetime action.}
\begin{equation} \label{eq:ztc}
    D^{}_{[\mu} \tau^{}_{\nu]}{}^A = 0\,
    \quad\implies\quad
    d\tau^A = - \Omega^A{}_B \wedge \tau^B.
\end{equation}
While part of the components in \eqref{eq:ztc} can be used to solve for the longitudinal spin connection~$\Omega_\mu{}^{AB}$, the rest give rise to the geometric constraints,
\begin{equation} \label{eq:goczt}
    E^\mu{}^{\phantom{(}}_{\!A'} \, E^\nu{}^{\phantom{(}}_{\!B'} \, \p^{\phantom{(}}_{[\mu} \tau^{\phantom{(}}_{\nu]}{}^A = 0\,,
        \qquad%
    E^\mu{}^{\phantom{(}}_{\!A'} \, \tau^\nu{}^{\phantom{(}}_{(\!A} \, \eta^{\phantom{(}}_{B)C} \, \p^{\phantom{(}}_{[\mu} \tau^{\phantom{(}}_{\nu]}{}^C = 0\,.
\end{equation}
The first condition in \eqref{eq:goczt} coincides with the integrability condition \eqref{eq:icp2} in the codimension-two foliation structure, which sets the torsion associated to $d\tau^A$ to zero.
The target-space geometry coupled to NR strings described by the worldsheet action~\eqref{eq:actionhb} with string Bargmann symmetries is referred to as \emph{string Newton-Cartan} (SNC) geometry, which has zero torsion~\cite{Andringa:2012uz}.
As desired, it is shown in \cite{Yan:2019xsf} that this action does not receive quantum correction to the $\lambda\bar{\lambda}$ operator at all loops. Therefore, the quantum sigma model \eqref{eq:actionhb} is renormalizable when the string Bargmann symmetries are imposed.
It therefore gives rise to a notion of NR string theory defined by a renormalizable quantum sigma model, which can be studied on its own in a self-consistent way, without referring to the full relativistic string theory.

The beta-functionals of the background fields $H_{\mu\nu}$\,, $B_{\mu\nu}$\,, and $\Phi$ in~\eqref{eq:actionhb} have been studied in~\cite{Gomis:2019zyu, Yan:2019xsf}, which we review below. In the path integral, the St\"{u}ckelberg symmetries \eqref{eq:Stueckelberg} that mix $H_{\mu\nu}$ and $B_{\mu\nu}$ are recast in the form of Ward identities. The physical beta-functionals are therefore manifestly invariant under the St\"{u}ckelberg symmetries in \eqref{eq:Stueckelberg}.
It has been proven in \cite{Yan:2019xsf} that there is a nonrenormalization theorem for $\tau_\mu{}^A$\,.
We can therefore define the beta-functionals with respect to the remaining variables $E_{\mu\nu}$\,, $M_{\mu\nu}$\,, and $\Phi$ that enter in the action \eqref{eq:cbsaEM} where the St\"{u}ckelberg symmetries are fixed,
\begin{equation}
    \beta^{(E)}_{\mu\nu} = \frac{dE_{\mu\nu}}{dt}\,,
        \qquad%
    \beta^{(M)}_{\mu\nu} = \frac{dM_{\mu\nu}}{dt}\,,
        \qquad%
    \beta^{(\Phi)} = \frac{d\Phi}{dt}\,.
\end{equation}
Here, $t$ is the renormalization `time' and $e^t$ defines the renormalization scale. In terms of the variables $H_{\mu\nu}$ and $B_{\mu\nu}$\,, we have
\begin{subequations}
\begin{align}
    \beta^{(E)}_{A'B'} & = \beta^{(H)}_{A'B'},
        &
    \beta^{(M)}_{AB} & = - \tfrac{1}{2} \, \epsilon_{AB} \left( \eta^{CD} \, \beta^{(H)}_{CD} -  \epsilon^{CD} \, \beta^{(B)}_{CD} \right), \\[2pt]
    \beta^{(M)}_{A'B'} & = \beta^{(B)}_{A'B'}\,,
        &
    \beta^{(M)}_{AA'} & = \beta^{(B)}_{AA'} + \epsilon_A{}^B \, \beta^{(B)}_{BA'}\,.
\end{align}
\end{subequations}
The subscripts $A$ and $A'$ of the beta-functionals denote contractions with the inverse vielbeine $\tau^\mu{}_A$ and $E^\mu{}_{A'}$, respectively.
In order to present these results, we first review some additional elements of SNC geometry. The geometric quantities that are invariant under the string Galilei boosts are $E^\mu{}_{A'}$\,, $\tau_\mu{}^A$\,, $H_{\mu\nu}$\,, and
\begin{equation}
    N^{\mu\nu} =
    \eta^{AB} \, \tau^\mu{}_A \, \tau^\nu{}_B - 2 \,  E^{(\mu}{}_{A'} \, \tau^{\nu)}{}_{A} \, m_\lambda{}^A \, E^\lambda{}_{A'}.
\end{equation}
A compatible connection can be constructed using these boost-invariant objects (but not uniquely, see for example~\cite{Bergshoeff:2019pij}),
\begin{equation}
    \Gamma^\rho{}_{\mu\nu} = \tfrac{1}{2} \, N^{\rho\sigma}  \left( \p_\mu \tau_{\nu\sigma} + \p_\nu \tau_{\mu\sigma} - \p_\sigma \tau_{\mu\nu} \right) + \tfrac{1}{2} \, E^{\rho\sigma} \left( \p_\mu H_{\nu\sigma} + \p_\nu H_{\mu\sigma} - \p_\sigma H_{\mu\nu} \right).
\end{equation}
Using this connection, the Riemann curvature $R^\rho{}_{\mu\nu\sigma}$\,, the Ricci tensor $R_{\mu\nu}$\,, and the covariant derivative $\nabla_{\!\mu}$ can be defined in the standard way. The beta-functionals are then given by
\begin{subequations}
\begin{align}
    \beta^E_{\mu\nu} & = \alpha' \, E_\mu{}^{A'} E_\nu{}^{B'} P_{A'B'} + O(\alpha'{}^2)\,,
        &%
    \beta^M_{AB} & = - \tfrac{\alpha'}{2} \, \epsilon_{AB} \left( P^C{}_D - \epsilon^{CD} \, Q_{CD} \right) + O(\alpha'{}^2)\,, \\[2pt]
    \beta^M_{\mu\nu} & = \alpha' \, Q_{A'B'} + O(\alpha'{}^2)\,,
        &%
    \beta^M_{AA'} & = \alpha' \left( P_{AA'} + \epsilon_A{}^B \, Q_{BA'} \right) + O(\alpha'{}^2)\,,
\end{align}
\end{subequations}
where
\begin{subequations}
\begin{align}
    P_{\mu\nu} & = R_{\mu\nu} + 2 \, \nabla_{\!\mu} \nabla_{\!\nu} \Phi - \tfrac{1}{4} \, E^{\rho\sigma} \, E^{\kappa\lambda} \, \mathcal{H}_{\mu \rho\kappa} \, \mathcal{H}_{\nu \sigma\lambda}\,, \\[4pt]
    Q_{\mu\nu} & = - E^{\rho\sigma} \left( \tfrac{1}{2} \nabla_{\!\rho} \mathcal{H}_{\sigma\mu\nu} - \nabla_{\!\sigma} \Phi \, \mathcal{H}_{\sigma\mu\nu} \right).
\end{align}
\end{subequations}
We denoted the Kalb-Ramond field strength as $\mathcal{H} = dB$\,.
The beta-functional associated with the dilaton is
\begin{align}
\begin{split}
    & \quad \beta^\Phi - \tfrac{1}{2} \, \beta(\ln G) \\[2pt]
    & = \frac{d\!-\!26}{6} - \alpha' \, E^{\mu\nu} \Bigl( \nabla_{\!\mu} \nabla_{\!\nu} \Phi - \nabla_{\!\mu} \Phi \, \nabla_{\!\nu} \Phi + \tfrac{1}{4} \, R_{\mu\nu} - \tfrac{1}{48} \, \mathcal{H}_{\mu A'B'} \, \mathcal{H}_{\nu}{}^{A'B'} \Bigr) + O(\alpha'{}^2)\,.
\end{split}
\end{align}
Here, $G = \det (\tau_\mu{}^A, E_\mu{}^{A'})$\,.
These one-loop beta-functionals have been derived first from analyzing the OPEs in \cite{Gomis:2019zyu}, and they were later corroborated by the background field method in~\cite{Yan:2019xsf}. The same result was also derived as a NR limit of the relativistic beta-functionals in~\cite{Bergshoeff:2019pij}.
The overlap of these results with the beta-functionals derived in~\cite{Gallegos:2019icg} for the sigma models in torsional Newton--Cartan geometry from \S\ref{ssec:null-red} is confirmed using double field theory methods in~\cite{Gallegos:2020egk}.

In \S\ref{sec:smbf}, we noted that the sigma model \eqref{eq:actionhb} is also invariant under a dilatational symmetry, namely,
\begin{equation} \label{eq:dst}
    \Phi \rightarrow \Phi + \ln \Delta\,,
        \qquad%
    \tau_\mu{}^A \rightarrow \Delta \, \tau_\mu{}^A\,,
        \qquad%
    \lambda \rightarrow \Delta^{-1} \, \lambda\,,
        \qquad%
    \bar{\lambda} \rightarrow \Delta^{-1} \, \bar{\lambda}\,.
\end{equation}
Nevertheless, this dilatational symmetry is not preserved by the zero-torsion constraint \eqref{eq:ztc}, unless $\p_{A'} \Delta = 0$ \cite{Bergshoeff:2019pij}. Intriguingly, one can still obtain a renormalizable worldsheet QFT that is compatible with the dilatational symmetry if one breaks half of the $Z_A$ symmetry~\cite{Yan:2021lbe}. In order to preserve the longitudinal Lorentz symmetry, the only possibility is to break a lightlike component of the $Z_A$ symmetry. We choose to break $\bar{Z} = Z_0 - Z_1$ for concreteness. Taking a contraction in the string Bargmann algebra that decouples the generator $\bar{Z}$ leads to a self-consistent subalgebra.
Gauging this algebra gives rise to the same string Galilei boost transformations, but now $m_\mu{}^A$ only transforms under $Z= Z_0 +Z_1$ as far as the extended symmetries are concerned, with
\begin{equation} \label{eq:hdm}
    \delta m_\mu{}^0 = \delta m_\mu{}^1 = D_\mu \sigma\,,
\end{equation}
where $\sigma$ is the Lie group parameter associated with the $Z$ generator. Requiring that \eqref{eq:actionhb} is invariant under the $Z$ symmetry leads us to the condition
\begin{equation} \label{eq:Dt}
    D^{}_{[\mu} \bar{\tau}^{}_{\nu]} = 0\,.
\end{equation}
This condition leads to the geometric constraints
\begin{equation} \label{eq:hgc}
    E^\mu{}_{A'} \, E^\nu{}_{B'} \, \p_{[\mu} \bar{\tau}_{\nu]} = 0\,,
        \qquad%
    E^\mu{}_{A'} \, \tau^\nu \, \p_{[\mu} \bar{\tau}_{\nu]} = 0\,.
\end{equation}
Here, $\tau^\mu = \frac{1}{2} \bigl( \tau^\mu{}_0 \, + \, \tau^\mu{}_1 \bigr)$.
These are the same constraints we encountered in~\eqref{eq:hztcs} in the context of the NR limit of supergravity.
Both constraints in~\eqref{eq:hgc} preserve the dilatational symmetry~\cite{Bergshoeff:2021tfn}, which is generated by the transformations in \eqref{eq:dst}. Moreover, only half of the integrability conditions \eqref{eq:icp2} appear, coinciding with the first condition in \eqref{eq:hgc}.

This halved $Z_A$ symmetry leads to attractive features. First, it is shown in \cite{Yan:2021lbe} that the geometric constraints \eqref{eq:hgc} are sufficient for eliminating at all loops the quantum corrections that generate the $\lambda\bar{\lambda}$ operator, which could
otherwise deform the theory towards relativistic string theory. Therefore, the symmetry algebra with a halved $Z_A$ symmetry still leads to a self-contained notion of NR string theory. Further analysis of the RG structure of the associate sigma model at the lowest order in $\alpha'$ has been explored in \cite{Yan:2021lbe}.
Second, it was suggested in~\cite{Harmark:2018cdl,Harmark:2019upf, Bergshoeff:2021bmc, Bergshoeff:2021tfn} that the original zero-torsion constraint \eqref{eq:ztc} might be too strong.
As discussed in \cite{Harmark:2018cdl,Harmark:2019upf}, torsion in $\tau_\mu{}^A$ seems to be essential for applications of NR strings to the AdS/CFT correspondence (see \S\ref{sec:nhd}).
Additionally, as we reviewed in \S\ref{eq:tsgl}, it was shown in~\cite{Bergshoeff:2021tfn} that the finiteness of the supersymmetry rules under the NR limit leads to a set of weaker torsion constraints, which coincide with the ones in \eqref{eq:hgc}.
While the string Bargmann proposal for the modified symmetry algebra requires the zero-torsion condition~\eqref{eq:ztc}, the modified proposal with a halved $Z_A$ symmetry still allows half of the $\tau_\mu{}^A$ to be torsional.
This modified proposal therefore probes a larger class of target-space gravities, which might eventually be useful for studies of NR supergravity and gauge-gravity duality.

\subsection{Nonrelativistic open strings and DBI action}
We already learned in \S\ref{sec:nrosncos} that, in flat spacetime, open strings ending on a D($d-2$)-brane that is transverse to the longitudinal spatial direction $X^1$ enjoy a Galilean-invariant dispersion relation. In a curved spacetime, we consider the boundary condition
$X^\mu \big|_{\p\Sigma} = f^\mu (Y^i)$\,,
where $Y^i$ parametrize the D-brane submanifold, with $i = 0, 2, \dots, d-1$\,, and $f^\mu$ describes how the D-brane is embedded in the target space. The Dirichlet boundary condition $\delta X^1\big|_{\p\Sigma} = 0$ in flat spacetime now becomes
$\delta X^\mu \big|_{\p\Sigma} = \delta Y^i \, \p_i f^\mu (Y^j)$\,.
In addition to the closed string sigma model~\eqref{eq:actionhb}, we now have an additional boundary action,
\begin{equation} \label{eq:Sbdry}
    S_\text{bdry} = \frac{1}{2\pi\alpha'} \int_{\p\Sigma} d\sigma^2 \, \left[ N(Y) \, \bigl( \lambda - \bar{\lambda} \bigr) + i \, A_i \, \p_\tau Y^i \right].
\end{equation}
Varying $\lambda$ and $\bar{\lambda}$ in this action sets $N = 0$\,, which means that $N$ decouples in any analysis of the open string action. However, it is important to note that there will be counterterms generated for $N$ once quantum corrections are turned on; this will give rise to a nontrivial beta-functional for $N$\,. Assuming that $X^1$ is an isometry direction, then setting $N = 0$ classically means that we are in the unbroken phase of the spontaneous breaking of the translational symmetry in $X^1$ due to the presence of the D-brane. The background field $N$ is associated to the Nambu-Goldstone (NG) boson that perturbs the shape of the D-brane. This NG boson is absorbed into the definition of the collective coordinate $Y^i$ in the case where $N = 0$\,. This can be seen by applying the T-duality rules \eqref{eq:Tduallambda} in flat space, where they map $\lambda - \bar{\lambda}$ to $i \, \p _\tau Y^1$, so that $N$ becomes a component of the gauge boson on the D-brane in the dual picture.

The beta-functionals of $N$ and $A_i$ have been derived in \cite{Gomis:2020fui} using the covariant background field method. Imposing Weyl invariance at the quantum level requires that these beta-functionals vanish.
These are target-space equations that arise from a D$(d-2)$-brane action, which has a straightforward generalization to D$p$-branes,
\begin{equation} \label{eq:nrsp}
    S_{\text{D}p} = - T_{p} \int d^{p+1} Y \, e^{-\Phi} \sqrt{- \det
    \begin{pmatrix}
        0 & \,\, \tau_\nu \, \p_j f^\nu \\[4pt]
        \tau_\mu \, \p_i f^\mu & \,\, \left( H_{\mu\nu} + B_{\mu\nu} \right) \p_i f^\mu \, \p_j f^\nu + F_{ij}
    \end{pmatrix}}\,,
\end{equation}
where $F = dA$ is the field strength on the D$p$-brane. A related worldvolume action for D-branes also appears in \cite{Kluson:2019avy, Kluson:2020aoq}, where the embedding spacetime is taken to be torsional Newton-Cartan spacetime extended with a periodic space direction, as discussed in \S\ref{ssec:null-red}.
In the flat limit, at the quadratic order in the field strength $F_{\mu\nu}$\,, the D$p$-brane action \eqref{eq:nrsp} in the broken phase reproduces Galilean electrodynamics, which is the $U(1)$ case of~\eqref{eq:gym}, with the scalar $N$ receiving the natural interpretation as a NG boson.

The DBI-like action \eqref{eq:nrsp} continues to hold in the case that the D-brane extends in the compactified longitudinal $X^1$ direction. In the flat limit, open strings residing on such a D-brane configuration satisfy the Neumann boundary condition in the $X^1$ circle. As reviewed in \S\ref{sec:nrosncos}, such open strings are in the NCOS sector and enjoy a relativistic dispersion relation.
It is shown in \cite{Gomis:2020izd} that dualizing the DBI-like action \eqref{eq:nrsp} that describes a D-brane localized in a longitudinal lightlike circle gives rise to the DLCQ of NCOS.
This is in contrast to the fact that the T-dual of a NR D$p$-brane localized in a longitudinal spatial circle is T-dual to relativistic DBI action in the DLCQ \cite{Gomis:2020izd}.

\subsection{Generalized nonrelativistic \texorpdfstring{$p$}{p}-branes} \label{sec:pbl}

In \cite{Gomis:2020fui}, it is shown that the Galilean DBI action \eqref{eq:nrsp} arises as the $c \rightarrow \infty$ limit of the relativistic DBI action,
\begin{equation}
    \hat{S}_\text{DBI} = - T_p \int d^{p+1} Y \, e^{-\hat{\Phi}} \sqrt{- \det
    \Bigl[
        \left( \hat{G}_{\mu\nu} + \hat{B}_{\mu\nu} \right) \p_i f^\mu \, \p_j f^\nu + \hat{F}_{ij}
    \Bigr]}\,,
\end{equation}
where the background fields are parametrized as in \eqref{eq:gbexp}, and $\hat{F}_{ij} = F_{ij}$ is $c$-independent.
This clarifies how higher-dimensional objects such as the D$p$-branes fit into the framework of NR string theory, which arises as a `stringy' limit of relativistic string theory.
Such a stringy limit induces a two-dimensional foliation in the spacetime geometry.
In contrast, generalizations of this stringy limit to the so-called `$p$-brane' limits have been discussed in the literature \cite{Gomis:2000bd, Brugues:2004an, Brugues:2006yd, Roychowdhury:2019qmp, Pereniguez:2019eoq, Kluson:2020rij}. A $p$-brane limit is usually applied to the Nambu-Goto action of relativistic $p$-branes coupled to a $(p+1)$-form gauge field $\hat{A}^{(p+1)}$. The relativistic $p$-brane action is
\begin{align} \label{eq:spbrel}
    \hat{S}_{p\text{-brane}} = - \int d^{p+1} Y \sqrt{-\det \left( \p_\alpha X^\mu \, \p_\beta X^\nu \, \hat{G}_{\mu\nu} \right)}
    - \int \hat{A}^{(p+1)}\,.
\end{align}
Instead of the `stringy' parametrization \eqref{eq:gbexp}, we now consider a `$p$-brane' parametrization,
\begin{equation}
    \hat{G}^{}_{\mu\nu} = c^2 \, \gamma^{}_{\mu}{}^u \, \gamma^{}_\nu{}^v + c^{1-p} \, E^{}_{\mu\nu}\,,
        \qquad%
    \hat{A}^{(p+1)}_{\mu_0 \cdots \mu_p} = - c^{p+1} \, \gamma^{}_{\mu_0}{}^{u_0} \cdots \gamma^{}_{\mu_p}{}^{u_p} \, \epsilon^{}_{u_0 \cdots u_p} + A^{(p+1)}_{\mu_0 \cdots \mu_p}\,.
\end{equation}
Here, $u = 0, \cdots, p$\,, and $\gamma_\mu{}^u$ are the vielbein fields that encode the geometry of the induced $(p+1)$-dimensional foliation in spacetime. In the limit $c \rightarrow \infty$\,, the $p$-brane action \eqref{eq:spbrel} gives rise to a nonsingular low-energy action,
\begin{equation} \label{eq:nrspb}
    S_{p\text{-brane}} = - \frac{1}{2} \int d^{p+1} \sigma \sqrt{-\gamma} \, \gamma^{\alpha\beta} \, \p_\alpha X^\mu \, \p_\beta X^\nu \, E_{\mu\nu} - \int A^{(p+1)},
\end{equation}
where $\gamma_{\alpha\beta} = \p_\alpha X^\mu \, \p_\beta X^\nu \, \gamma_{\mu\nu}$ and $\gamma = \det \gamma_{\alpha\beta}$\,. In the case of $p = 1$\,, the $p$-brane action \eqref{eq:nrspb} is the Nambu-Goto action \eqref{eq:nrngf} of NR string theory. For the Polyakov description of the $p$-brane limit, see, e.g., \cite{Gomis:2004pw, Gomis:2005bj}, where $\kappa$-symmetries of the $p$-brane actions are also discussed. Also see \cite{Gomis:2005pg} for discussions on $\kappa$-symmetry in the Green-Schwarz formalism of NR string theory in AdS background.

The $p$-brane limit of relativistic $p$-branes differs in nature from the stringy limit of relativistic DBI action. The NR DBI action \eqref{eq:nrsp} that arises as the stringy limit describes D$p$-branes coupled to SNC geometry, equipped with a two-dimensional foliation structure. In contrast, the NR $p$-branes described by \eqref{eq:nrspb} are coupled to the so-called $p$-brane Newton-Cartan geometry, equipped with a $(p+1)$-dimensional foliation structure. The two-brane limit that induces a three-dimensional foliation structure has been applied to (super) M2-branes~\cite{Gomis:2004pw, Kluson:2019uza}.
Also see \cite{Gopakumar:2000ep, Bergshoeff:2000ai} for a theory of light Open Membranes (OM) on an M5-brane near a critical three-form field strength. This OM theory arises in the two-brane limit of M-theory and describes five-dimensional NCOS at strong coupling. Such an (open) membrane limit of M2- and M5-brane actions has been studied in \cite{Garcia:2002fa}.
Moreover, the two-brane limit of the bosonic sector of eleven-dimensional supergravity has recently been investigated in \cite{Blair:2021waq}. S-dualities of light (super) D$p$-branes that arise from performing the general $p$-brane limits in relativistic string/M-theory are studied in \cite{Gomis:2000bd, Kamimura:2005rz}.

\section{Nonrelativistic Holographic Dualities} \label{sec:nhd}
As was mentioned in~\S\ref{sec:intro}, one of the hopes in studying NR string theory is that it will allow us to obtain a better understanding of previously inaccessible corners of relativistic string theory.
We will now briefly discuss particular applications of these ideas in the context of the AdS/CFT correspondence.
Other directions will be mentioned in~\S\ref{sec:co}.

Specifically, we focus on a decoupling limit that was originally introduced in the context of $\mN=4$ super-Yang--Mills, which is known as the Spin Matrix theory (SMT) limit~\cite{Harmark:2014mpa}.
We first introduce this limit by comparison with the Penrose/BMN limit~\cite{Blau:2002dy,Berenstein:2002jq}.
Subsequently, we discuss the associated worldsheet theory as well as the backgrounds and sigma models that result from applying the SMT limit to strings on AdS$_5\times S^5$.

\subsection{Decoupling limits of the AdS/CFT correspondence}
In the strongest form of the AdS/CFT correspondence~\cite{Maldacena:1997re}, four-dimensional $\mN=4$ supersymmetric Yang--Mills theory is conjectured to capture the full nonperturbative dynamics of IIB string theory on AdS${}_5 \times S^5$.
However, even in the large $N$ limit, matching perturbative string theory results explicitly to the dual field theory is a formidable task.
For this reason, various limits that focus on a simpler decoupled sector have been considered.
One example is the well-known BMN limit~\cite{Berenstein:2002jq}, which is related to the Penrose limit~\cite{Blau:2002dy} of AdS${}_5 \times S^5$, zooming in on the neighborhood of a lightlike geodesic.
In terms of the background geometry, this limit can be obtained by boosting along an equator of $S^5$ at the center of AdS${}_5$\,, leading to an infinite angular momentum $J$ around the equator.
The resulting geometry is a ten-dimensional pp-wave~\cite{Blau:2001ne}, a maximally supersymmetric IIB solution.
On the dual gauge theory side, this limit selects the BMN states that carry infinite R-charge $J$\,, with
\begin{equation}
    Q \rightarrow \infty\,,
        \qquad%
    \frac{\Delta - Q}{Q} \rightarrow 0\,.
\end{equation}
Here, $\Delta$ is the anomalous dimension of the corresponding operator, and $Q$ is a particular combination of the R-charge and $S^3$ Cartan generators $(J_1,J_2,J_3)$ and $(S_1,S_2)$, which is usually taken to be equal to $J_1$.
Other choices of $Q$ would lead to different coordinates on the pp-wave.
In this limit, the field theory spectrum can be matched perturbatively to the spectrum obtained from quantizing the string on the pp-wave background.

NR string theory can be viewed as another example where a decoupled sector of string theory is analysed in a self-contained way.
In this case, the decoupled sector only contains winding string states that satisfy a Galilean-invariant dispersion relation.
As a first example, an NR limit of string theory on AdS${}_5 \times S^5$ has been studied in~\cite{Gomis:2005pg}, where it was shown to be equivalent to a supersymmetric two-dimensional sigma model of free particles propagating on~AdS$_{2}$\,.
The T-dual of this NR string theory leads to relativistic strings on a time-dependent pp-wave with a compactified lightlike circle and hence DLCQ.
This NR string theory was subsequently given a dual interpretation in terms of a conformal quantum mechanics theory~\cite{Sakaguchi:2007ba}.
See also~\cite{Fontanella:2021hcb,Fontanella:2021btt} for more recent work in this direction.

Another decoupling limit that allows us to probe particular subsectors of $\mN=4$ SYM on $\RR\times S^3$ is known as \emph{Spin Matrix theory} (SMT)~\cite{Harmark:2014mpa}.
At least in principle, this limit is tractable at finite values of $N$, which would allow it to capture perturbative and nonperturbative gravity effects in AdS.
In this case, nonrelativistic behavior arises by zooming in on the dynamics of the theory close to a BPS bound.
Given a BPS bound $\Delta \geq Q$ of $\mN=4$, the associated SMT is defined by the limit
\begin{equation}
\label{eq:smt-ft-limit}
    \lambda \to 0\,,
    \qquad
    N = \text{fixed},
    \qquad
    \frac{\Delta-Q}{\lambda} = \text{fixed}.
\end{equation}
Here, $Q$ is again a particular combination of $S^3$ and R-charge Cartan generators.
Fields in SMT have `matrix' indices that are inherited from the $SU(N)$ gauge symmetry of the parent $\mN=4$ theory.
Similarly, they have a `spin' index, determined by the choice of $Q$, corresponding to the spin symmetry group of the remaining states.
In the $N\to\infty$ limit, SMT reduces to a spin chain of length equal to the total R-charge $J$, so that $J\to\infty$ gives a continuum limit.
Subleading $1/N$ corrections then enable the splitting and joining of the spin chains.
The simplest nontrivial choice corresponds to $Q=J_1 + J_2$, which leads to a Heisenberg spin chain for $N\to\infty$ and to the $SU(2)$ Landau--Lifshitz model in the continuum limit.
The largest SMT corresponds to $Q=J_1+J_2+J_3+S_1+S_2$ and results in the spin group $PSU(1,2|3)$, where the 1/16 BPS supersymmetric black hole in AdS${}_5 \times S^5$ from~\cite{Gutowski:2004yv} survives in the limit.

The field theory SMT limit~\eqref{eq:smt-ft-limit} has been mapped to the relativistic string sigma model on AdS$_5\times S^5$~\cite{Harmark:2017rpg}.
It results in a NR string similar to the ones we discussed before, but in this case not only the target space geometry but also the worldsheet geometry is nonrelativistic.
As a result, instead of the usual Virasoro symmetry, this Spin Matrix string theory has a Galilean conformal algebra (GCA) of reparametrization symmetries on the worldsheet~\cite{Harmark:2018cdl}.
Its target-space geometry, known as $U(1)$-Galilean geometry, is closely related to the TSNC geometry that we have discussed in \S\ref{sec:tsncgfal}, and we will discuss its local symmetries below.

In the following, we give a brief review of the construction of the sigma model for SMT strings from the sigma model for NR strings in general backgrounds.
Additionally, we show how the $U(1)$-Galilean geometries associated to particular BPS bounds can be constructed from an appropriate limit of the AdS geometry~\cite{Harmark:2017rpg,Harmark:2018cdl,Harmark:2020vll}.
We also review how, after gauge fixing the GCA worldsheet symmetry, the classical string sigma model reproduces known effective continuum spin chain actions obtained from field theory.
Finally, we discuss their interaction with the Penrose limits mentioned above.

\subsection{Spin Matrix theory limit of nonrelativistic string theory}
We first consider the tensionless limit of the NR string sigma model~\eqref{eq:cbsaEM} that was introduced in~\cite{Harmark:2017rpg,Harmark:2018cdl}.
This limit results in a NR string with a NR worldsheet structure.
For simplicity, we will set the dilaton to zero in the following discussion.
Additionally, following the discussion in \S\ref{ssec:null-red}, we assume the existence of a lightlike Killing vector in the original Lorentzian target space,
which results in a TSNC geometry with a longitudinal spatial isometry.
Correspondingly, we split the embedding coordinates as $X^\mu=(X^\hmu,\eta)$, where the worldsheet scalar~$\eta$ parametrizes the longitudinal spatial isometry, and $i$ is a $(d-1)$-dimensional index that contains a timelike component.
Here, following the literature, we now consider a Lorentzian worldsheet. Using the parametrization~\eqref{eq:tnc-snc-identification} together with $b_{\hmu}=0$ and $M_{\hmu\hnu}=0$, this leads to the worldsheet action
\begin{align} \label{eq:setnc}
    S = & - \frac{1}{4\pi\alpha'} \int_\Sigma d^2 \sigma \left( \sqrt{-h} \, h^{\alpha\beta} \, \p_\alpha X^\hmu \, \p_\beta X^\hnu \, E_{\hmu\hnu}
    + 2 \,  \, \epsilon^{\alpha\beta} \, \p_\alpha X^i m_i \, \p_\beta \, \eta \right) \\[2pt]\nonumber
    & - \frac{1}{4\pi\alpha'} \int_\Sigma d^2 \sigma  \Bigl[
    \lambda \, \epsilon^{\alpha\beta} \left(e_\alpha{}^0 + e_\alpha{}^1\right)\left( \pd_\beta X^i\tau^{}_i + \pd_\beta \eta\right)
    + \bar\lambda \,  \epsilon^{\alpha\beta} \left(e_\alpha{}^0 - e_\alpha{}^1\right)\left(\pd_\beta X^i \tau^{}_i - \pd_\beta \eta\right)
    \Bigr]\,.
\end{align}
We have introduced a set of worldsheet vielbeine $e_\alpha{}^a$ with $a = 0, 1$ such that $h_{\alpha\beta} = \eta_{ab} \, e_\alpha{}^a e_\beta{}^b$.
As discussed in \S\ref{ssec:null-red}, this sigma model describes strings propagating in a TNC geometry that is extended with a spatial circle direction parametrized by $\eta$.
The TNC geometry is described by the clock one-form $\tau^{}_\hmu$\,, the transverse vielbeine $E_{\hmu}{}^{A'}$, and the $U(1)$ gauge field~$m_\hmu$ corresponding to the remaining antisymmetric couplings.

Next, we perform a zero tension limit of the sigma model \eqref{eq:setnc}.
This is implemented by sending $c\to\infty$ after taking the following rescalings:
\begin{subequations} \label{eq:smtc}
\begin{align}
    \alpha' & \rightarrow c \, \alpha'\,,
        &%
    e_\alpha{}^1 & \rightarrow c \, e_\alpha{}^1\,,
        &%
    \lambda & \rightarrow \frac{\omega + c \, \psi}{2 \, c^3}\,,
     \\[2pt]
    \eta & \rightarrow c \, \eta\,,
        &%
    e_\alpha{}^0 & \rightarrow c^2 \, e_\alpha{}^0\,,
        &%
    \bar{\lambda} & \rightarrow \frac{\omega - c \, \psi}{2 \, c^3}\,,
        &%
    \tau^{}_i & \rightarrow c^2 \, \tau^{}_i\,,
\end{align}
\end{subequations}
while $E_i{}^{A'}$ and $m_i$ remain unchanged.
In this limit, the action~\eqref{eq:setnc} becomes
\begin{align} \label{eq:smtwt}
\begin{split}
    S & = - \frac{1}{4\pi\alpha'} \int d^2 \sigma \, \Bigl( e \, e^\alpha{}_1 \, e^\beta{}_1 \, \p_\alpha X^\hmu \, \p_\beta X^\hnu \, E_{\hmu\hnu} +2 \, \epsilon^{\alpha\beta} \, \p_\alpha X^\hmu \, m_\hmu \, \p_\beta \eta \Bigr) \\[2pt]
    & \quad - \frac{1}{4\pi\alpha'} \int d^2 \sigma \, \epsilon^{\alpha\beta} \Bigl[ \omega \, e_\alpha{}^0 \, \p_\beta X^\hmu \, \tau^{}_\hmu + \psi \bigl( \,
    e_\alpha{}^0 \, \p_\beta \eta
    + e_\alpha{}^1 \, \p_\beta X^\hmu \, \tau^{}_\hmu
    \bigr) \Bigr]\,.
\end{split}
\end{align}
Similar to the previous action, this sigma model is invariant under global worldsheet transformations corresponding to Galilean boosts $\Lambda_{A'}$ and $U(1)$ symmetries $\sigma$,
\begin{equation}
    \delta \tau_\hmu = 0,
    \qquad
    \delta m_\hmu = \pd_\hmu \sigma,
    \qquad
    \delta E_{\hmu\hnu} = 2 \, \tau_{(\hmu} \, E_{\hnu)}^{A'} \Lambda_{A'},
\end{equation}
which correspond to gauge symmetries in the target space geometry.
Unlike in TNC geometry, the $U(1)$ gauge field $m_\hmu$ field no longer transforms under Galilei boosts after the limit, which is why the resulting geometry is referred to as $U(1)$-Galilean geometry.
We can also clearly see that the worldsheet vielbeine $e_\alpha{}^0$ and $e_\alpha{}^1$ are treated differently in the action~\eqref{eq:smtwt}, which indicates the nonrelativistic structure on the worldsheet.
In fact, the sigma model is invariant under the local transformations
\begin{equation}
    e_\alpha{}^0 \rightarrow f \, e_\alpha{}^0\,,
        \qquad%
    e_\alpha{}^1 \rightarrow f \, e_\alpha{}^1 + g \, e_\alpha{}^0\,,
        \qquad%
    \omega \rightarrow f^{-1} \, \omega - g \, f^{-2} \, \psi\,,
        \qquad%
    \psi \rightarrow f^{-1} \, \psi\,,
\end{equation}
where $f$ parametrizes Weyl transformations and $g$ corresponds to local Galilei boosts.
We can choose flat gauge on the worldsheet to fix these local symmetries, up to the residual gauge transformations
\begin{equation}
    \sigma^0 \rightarrow F(\sigma^0)\,,
        \qquad%
    \sigma^1 \rightarrow F' (\sigma^0) \, \sigma^1 + G(\sigma^0)\,,
\end{equation}
which correspond to a \emph{Galilei conformal algebra}.
Classically, its Lie brackets are given by
\begin{equation}
    [L_n,\, L_m] = (n-m) \, L_{n+m}\,,
        \qquad%
    [L_n\,, M_m] = (n-m) \, M_{n+m}\,.
\end{equation}
This algebra corresponds to a contraction of the usual two Virasoro algebras.
\subsection{Spin Matrix strings from a limit on \texorpdfstring{AdS$_5\times S^5$}{AdS5xS5}}
We now consider the SMT limit of strings in AdS${}_5 \times S^5$~\cite{Harmark:2017rpg}, which can be implemented using the previous tensionless limit.
If we denote the AdS$_5$ and $S^5$ radius by $\ell$, the associated effective string tension is given by
\begin{equation}
    T = \frac{\ell^2}{2\pi\ell_s^2} = \sqrt{\frac{g_s \, N}{2\pi}}\,,
        \qquad%
    \ell_s^2 = 2 \alpha'\,,
\end{equation}
where $g_s$ is the string coupling and $N$ denotes the total number of coinciding extremal black D3-branes.
Recall that the AdS/CFT dictionary relates field theory and string theory parameters through
$4\pi\, g_s = \lambda/N$ and
$\ell/\ell_s = \lambda^{1/4}$,
where $\lambda$ is the 't Hooft coupling in $\mathcal{N} = 4$ supersymmetric Yang--Mills theory.
The SMT field theory limit~\eqref{eq:smt-ft-limit} then corresponds to
\begin{equation} \label{eq:smtlIIb}
    T \rightarrow 0\,,
        \qquad%
    N = \text{fixed},
    \qquad
    \frac{\Delta - Q}{T^2} = \text{fixed}.
\end{equation}
The energy $\Delta$ and the charge $Q$ are now associated to target space isometries of the AdS$_5\times S^5$ metric, which, in global coordinates, is given by
\begin{equation} \label{eq:adsm}
    ds^2 = \ell^2 \left( - \cosh^2 \! \rho \, dt^2 + d\rho^2 + \sinh^2 \! \rho \, d\Omega_3^2 + d\Omega_5^2 \right),
\end{equation}
where $d\Omega_k^2$ is the metric of a $k$-sphere of unit radius.
We now have $\Delta = i\pd_t$, and the choice of $Q = S + J$ corresponds to the choice of a particular combination $S$ of the Cartan generators of the $S^3$ and the $S^5$ isometries $J$.
We will illustrate this with concrete examples in the following.
Next, one introduces coordinates $x^0$ and $u$ such that~\cite{Harmark:2020vll}
\begin{equation}
    i \pd_{x^0}
    = \Delta - Q
    = \Delta - S - J\,,
    \qquad
    -i \pd_u = \frac{1}{2} \left(\Delta - S + J\right).
\end{equation}
The $u$ coordinate will be lightlike on a particular submanifold $M$ of AdS$_5\times S^5$ and, as we will see in more detail below, the SMT limit restricts the dynamics of the string to this submanifold.
We can parametrize the metric on $M$ as
\begin{equation}
    \left.ds^2 / \ell^2 \right|_{M}
    = 2 \, \tau_\hmu \, dx^\hmu \left( du + m_\hnu dx^\hnu \right) + E_{\hmu\hnu} \, dx^\hmu dx^\hnu.
\end{equation}
From this parametrization, we obtain a TNC geometry $(\tau_\hmu,m_\hmu, E_{\hmu\hnu})$, which allows us to construct the nonrelativistic bosonic TNC string action in \eqref{eq:setnc}.
Note that the dimension of this geometry will depend on the choice of $Q$.
In this construction, the momentum along the $u$-direction is interpreted as a winding number in the $\eta$ direction, as discussed in~\S\ref{ssec:null-red}.
Additionally, it turns out that the clock one-form $\tau_\hmu \, dx^\hmu = dx^0 + \cdots$ always contains a component along the $X^0$ direction.

The SMT field theory limit~\eqref{eq:smtlIIb} can then be implemented using the tensionless string limit.
For this, we scale the conserved charge associated to the $X^0$ direction as $T^2$,
\begin{equation}
  \label{eq:smt-limit-coord-scaling}
    c\to\infty,
    \qquad
    X^0 = c^2 \tilde{X}^0,
    \quad
    c = \frac{1}{2\sqrt{2}\pi T},
    \quad
    N \text{ and } \tilde{X}^0 \text{ fixed }.
\end{equation}
Note that, in this limit, the momentum along the $u$ direction (and hence the winding along the $\eta$ direction) is given by $J$, which corresponds to the length of the spin chain in the dual field theory.

Going to flat worldsheet gauge, solving the constraints imposed by the Lagrange multipliers $\omega$ and $\psi$, and fixing the residual GCA transformations, the action~\eqref{eq:smtwt} reduces to
\begin{equation}
\label{eq:smt-gauge-fixed-action}
    S = - \frac{J}{2\pi} \int d^2\sigma \left( m_i \, \p_0 {X}^i + \frac{1}{2} \, E_{ij} \, \p_1 X^i \,\p_1 X^j \right).
\end{equation}
As we will see below, this allows us to recover several known sigma models arising from spin chains.
So far, this gauge fixing of the SMT string has only been implemented classically.

\subsubsection{The \texorpdfstring{$SU(2)$}{SU(2)} Spin Matrix string and the Landau--Lifshitz model}
Now let us illustrate the SMT string construction with the concrete example of $Q=J_1+J_2$, which allows us to zoom in the $SU(2)$ sector~\cite{Harmark:2017rpg,Harmark:2018cdl,Harmark:2020vll}.
From the bulk perspective, this charge involves two of the three commuting $S^5$ isometries, so we can parametrize it using Hopf coordinates on an $S^3\subset S^5$.
In these coordinates, the $S^5$ metric is given by
\begin{subequations}
\begin{align}
    d\Omega_5^2 & = d\alpha^2 + \sin^2 \! \alpha \, d\Omega'_3{}^2\, + \cos^2 \! \alpha \,d\beta^2 , \\[2pt]
    d\Omega'_3{}^2 & = \frac{1}{4} \Bigl( d\theta^2 + \sin^2 \! \theta \, d\phi^2 \Bigr) + \Bigl( d\gamma + \frac{1}{2} \, \cos \theta \, d\phi \Bigr)^2\,,
\end{align}
\end{subequations}
where $\gamma$ parametrizes the Hopf fiber, and $(\theta,\phi)$ parametrize the base $S^2$ of the $S^3\subset S^5$.
Correspondingly, we have $-i\pd_\gamma = J_1+J_2$.
Together with the global AdS time $t$ from~\eqref{eq:adsm}, we can then define the adapted coordinates $X^0$ and $u$,
\begin{equation}
    t = X^0 - \frac{u}{2}\,,
    \quad
    \gamma = X^0 + \frac{u}{2}
    \quad\implies\quad
    i \pd_{X^0} = \Delta - J_1 - J_2,
    \quad
    - i \, \pd_u = \frac{1}{2}\left( \Delta + J_1 + J_2 \right).
\end{equation}
The length of the $\pd_u$ vector is then given by
\begin{equation}
    \ell^{-2} \, \pd^2_u = - \cosh^2 \rho + \sin^2\alpha\,.
\end{equation}
Hence, we see that the $u$-direction is lightlike when $\rho=0$ and $\alpha=0$.
As a result, in the SMT limit zooming in on $Q = J_1 + J_2$, the dynamics of the string is restricted to the center of~AdS$_5$ and the Hopf $S^3$ inside $S^5$.
The $SU(2)$ spin group then arises from the isometries of this $S^3$.
The corresponding $U(1)$-Galilean background geometry is given by
\begin{equation}
    \label{eq:su2-u1-gal-geom}
    \tilde{\tau}_i \, dx^i = d\tilde{x}^0,
    \quad
    m_i \, dx^i = - \frac{1}{2} \cos\theta \, d\phi\,,
    \quad
    E_{ij} \, dx^i \, dx^j = \frac{1}{4} \left( d\theta^2 + \sin^2\theta \, d\phi^2 \right).
\end{equation}
On this background, the gauge fixed action~\eqref{eq:smt-gauge-fixed-action} corresponds to
\begin{equation}
\label{eq:smt-su2-ll-gauge-fixed-action}
    S = \frac{J}{4\pi} \int d^2 \sigma \left[
     \cos\theta \, \dot{\phi} - \frac{1}{4} \bigl( \theta'{}^2 + \sin^2 \theta \, \phi'{}^2 \bigr) \right],
\end{equation}
where $\dot{\phi} =  \p \phi / \p \sigma^0$ and $(\theta, \phi)' = \p (\theta, \phi) / \p \sigma^1$\,.
This is the Landau-Lifshitz sigma model that describes the XXX${}_{1/2}$ ferromagnetic Heisenberg spin chain in the large $J$ limit, with $\sigma^1$ the position on the spin chain, corresponding to the string winding in the periodic $\eta$ direction.
This sigma model has also been obtained from a similar limit of strings on AdS$_5\times S^5$ by Kruczenski~\cite{Kruczenski:2003gt}, although the latter limit involves taking $J\to\infty$ while keeping $\lambda/J^2$ fixed.
In contrast, the Spin Matrix limit corresponds to $\lambda\to0$ with $J$ fixed.

\subsubsection{General SMT string backgrounds and Penrose limits}
The most general SMT limit comes from zooming in on the BPS bound
\begin{equation}
    \Delta \geq Q = S_1 + S_2 + J_1 + J_2 + J_3\,,
\end{equation}
which involves all Cartan generators of the $S^3\subset\text{AdS}_5$ and the $S^5$ isometries.
The resulting $U(1)$-Galilean target-space geometry is given by
\begin{equation}
    \label{eq:psu123-u1-gal-background}
    \begin{gathered}
    \tilde{\tau}_i \, dx^i = d\tilde{x}^0,
    \qquad
    m_i \, dx^i = - \sinh^2 \! \rho \, dw - \sinh^2 \! \rho \, \bar{A} - B\,,
    \\[2pt]
    E_{ij} \, dx^i \, dx^j = d\rho^2 + \sinh^2\!\rho \, d\bar{\Sigma}_1^2
    + \sinh^2\! \rho \, \cosh^2\!\rho \left( dw + \bar{A}\right)^2 + d\Sigma_2^2\,.
    \end{gathered}
\end{equation}
Here, $d\Sigma_1^2$ and $d\Sigma_2^2$ refers to the Fubini--Study metric on $\mathbb{C}\mathbb{P}^1$ and $\mathbb{C}\mathbb{P}^2$, and $B$ and $A$ refer to their respective potentials,
\begin{equation}
    \begin{gathered}
    B = \sin^2\!\xi \left(d\psi + A \right) - \frac{1}{2} d\psi\,,
    \qquad
    A = \frac{1}{2} \cos\!\theta \, d\phi\,,
    \\[2pt]
    d\Sigma_2^2 = d\xi^2 + \sin^2\!\xi \,  d\Sigma_1^2 + \sin^2\!\xi \cos^2\!\xi \left(d\psi + A\right)^2,
    \qquad
    d\Sigma_1^2 = \frac{1}{4} \left( d\theta^2 + \sin^2\!\theta \, d\phi^2 \right).
    \end{gathered}
\end{equation}
They arise from the $S^5$, and from the $S^3$ in AdS$_5$ (corresponding to the barred quantities $\bar{A}$ and $d\bar{\Sigma}_1^2$).
The radial $\rho$ direction is inherited from AdS$_5$, and $w$ parametrizes the Hopf fiber of its $S^3$.
This geometry contains all other geometries corresponding to more restrictive BPS bounds.
In particular, we can recover the $SU(2)$ geometry in~\eqref{eq:su2-u1-gal-geom} by setting $\rho=0$ and $\xi=\pi/2$ as well as fixing $\psi$.
That restriction leaves us with the $(\theta,\phi)$ coordinates that parametrize the $SU(2)$ Landau-Lifshitz model in~\eqref{eq:smt-su2-ll-gauge-fixed-action}.
More generally, the $U(1)$-Galilean geometry associated to each SMT limit can be read off from Table~\ref{tab:smt-dof}.

\begin{table}[ht]
    \centering
    \begin{tabular}{r|c|c|c|c|c|c}
        {\small spin group} & {\small $Q$} & {\small $2n$} & {\small $\rho,w\in\text{AdS}_5$} & {\small $\bar\theta, \bar\phi \in\text{AdS}_5$}
        & {\small $\theta,\phi \in S^5$} & {\small $\xi,\psi\in S^5$}
        \\ \hline
        {\small $SU(2)$} & {\small $J_1 + J_2$} & 2 & -- & -- & \checkmark & --
        \\
        {\small $SU(2|3)$} & {\small $J_1 + J_2 + J_3$} & 4 & -- & -- & \checkmark & \checkmark
        \\
        {\small $SU(1,1)$} & {\small $S_1 + J_1$} & 2 & \checkmark & -- & -- & --
        \\
        {\small $PSU(1,1|2)$} & {\small $S_1 + J_1 + J_2$} & 4 & \checkmark & -- & \checkmark & --
        \\
        {\small $SU(1,2|2)$} & {\small $S_1 + S_2 + J_1$} & 4 & \checkmark & \checkmark & -- & --
        \\
        {\small $PSU(1,2|3)$} & {\small $S_1 + S_2 + J_1 + J_2 + J_3$} & 8 & \checkmark & \checkmark & \checkmark & \checkmark
    \end{tabular}
    \caption{The Spin Matrix theories arising from the near-BPS limit $\Delta\geq Q$ for charge combinations with integer coefficients, the $2n$ spatial dimensions of their target space in the bulk and their AdS$_5\times S^5$ origin.
    The associated $U(1)$-Galilean geometry can be obtained from appropriate restrictions of the $PSU(1,2|3)$ result in~\eqref{eq:psu123-u1-gal-background}.}
    \label{tab:smt-dof}
\end{table}

The resulting sigma models match with the $SU(1,1)$ sigma model and the bosonic part of the $PSU(1,1|2)$ sigma model that were obtained (following~\cite{Kruczenski:2003gt} for the $SU(2)$ sector) from a coherent state representation in~\cite{Stefanski:2004cw,Bellucci:2004qr,Bellucci:2006bv}.
Spinning string solutions of the SMT string sigma models have been considered in~\cite{Roychowdhury:2021wte}.

To simplify the resulting sigma models, we set $J\to\infty$\,, focusing on long spin chains or long strings, and hence zooming in on the region around a particular point of the $U(1)$-Galilean target space.
For example, in the $SU(2)$ Landau--Lifshitz sigma model~\eqref{eq:smt-su2-ll-gauge-fixed-action}, we can take
\begin{equation}
    J \to \infty\,,
    \qquad
    \theta = \frac{\pi}{2} + \frac{x}{\sqrt{{J}}}\,,
    \qquad
    \phi = \frac{y}{\sqrt{J}}\,,
\end{equation}
with $x$ and $y$ fixed.
Then the action~\eqref{eq:smt-su2-ll-gauge-fixed-action} becomes
\begin{equation}
    S = \frac{1}{4\pi} \int d^2\sigma \left( x \, \dot{y} - \frac{1}{2} \left[ (x')^2 + (y')^2 \right] \right),
\end{equation}
corresponding to the free magnon limit of the Landau--Lifshitz model.
This action can then be quantized exactly and its spectrum matches the result of the corresponding decoupling limit on the field theory side~\cite{Harmark:2008gm}.

Geometrically, the large charge limit is similar to the Penrose limit discussed above.
There, one zooms in on the neighborhood of a lightlike geodesic on AdS$_5\times S^5$, which results in a pp-wave geometry.
This is a maximally supersymmetric solution of IIB supergravity.
The solution is unique, but, depending on the choice of lightlike geodesic, one obtains the same geometry in different coordinates, analogous to how different $u$-coordinates correspond to different SMT limits.
The most general form of the pp-wave metric we need is
\begin{equation}
\label{eq:gen-pp-wave}
    ds^2 / \ell^2 = dx^0 \left( du + x_k \, dy^k \right) + \delta_{kl} \left( dx^k dx^l + dy^k dy^l \right)
    + \delta_{pq} \left[ dx^{p} dx^{q} - x^{p} x^{q} \bigl(dx^0\bigr)^2 \right].
\end{equation}
Here, $k,l=1,\ldots,n$ are `flat' directions, while the string feels a quadratic potential in the $p,q=1,\ldots,8-2n$ directions.
The SMT limit then suppresses the dynamics in the $p'$ directions, and results in the sigma model~\eqref{eq:smt-gauge-fixed-action} coupled to the $U(1)$-Galilean geometry
\begin{equation}
\label{eq:gen-ff-background}
    \tilde{\tau} = d\tilde{x}^0,
    \qquad
    m_i \, dX^i = - \sum_{k=1}^n x_k \, dy^k,
    \qquad
    E_{ij} \, dX^i \, dX^j = \sum_{a=1}^{n} \left[ (dx^k)^2 + (dy^k)^2 \right].
\end{equation}
The number of surviving spatial directions $2n$ is determined by the choice of SMT limit as listed in Table~\ref{tab:smt-dof}.
These $U(1)$-Galilean geometries were referred to as `flat-fluxed' or FF backgrounds in~\cite{Harmark:2020vll}, since they contain only the minimum requisite flux $m_i$ in order to make the resulting SMT sigma model~\eqref{eq:smtwt} nontrivial.

Supplemented with the appropriate five-form field strength, IIB strings can be quantized on the general pp-wave background~\eqref{eq:gen-pp-wave}, and the SMT limit of the resulting spectrum can be matched to the field theory result~\cite{Grignani:2009ny}.
To further establish this corner of NR strings in the holographic correspondence, this result should also be obtained from a direct quantization of the SMT string sigma model on the FF $U(1)$-Galilean backgrounds~\eqref{eq:gen-ff-background}.
Finally, a new approach has recently been developed for the explicit construction of Spin Matrix theories using a classical reduction of $\mathcal{N}=4$ followed by a suitable
 quantization and normal ordering procedure~\cite{Harmark:2019zkn,Baiguera:2020jgy,Baiguera:2020mgk,Baiguera:2021hky}.
This construction reproduces earlier results obtained~\cite{Harmark:2014mpa} from limits of the one-loop spectrum of $\mN=4$ and also leads to a two-dimensional field theory formulation of the $SU(1,1)$ sectors, suggesting a natural dual description for SMT strings on the three-dimensional $SU(1,1)$ geometry in Table~\ref{tab:smt-dof} at large $N$.

\section{Outlook} \label{sec:co}

We have reviewed recent developments in several aspects of NR string theory.
We discussed how this theory arises from a decoupling limit of relativistic string theory.
Starting from the free theory, we showed that this limit gives rise to a self-consistent, unitary and UV-complete string theory with a Galilean-invariant spectrum.
The resulting NR string theory provides a first-principles definition of the discrete light cone quantization (DLCQ) of relativistic strings that is covariant under nonrelativistic symmetries.
In general backgrounds, with appropriate symmetries imposed on the worldsheet theory,
NR string theory can be studied in a self-contained way, without referring to the parent relativistic string theory.
Several aspects of the geometry, the quantization and the dualities of this theory have already been explored, but much work remains to be done.

In particular, the supersymmetrization of NR string theory is still relatively unexplored.
In our discussion, we have exclusively discussed the bosonic sector of NR string theory, which has so far been the main focus of current research.
Nevertheless, there already have been studies of NR superstring analogs of the Ramond-Neveu-Schwarz~\cite{Kim:2007pc, Kim:2007hb, Blair:2019qwi}, the Green-Schwarz formalism~\cite{Gomis:2005pg, Park:2016sbw}, and $\kappa$-symmetries \cite{Gomis:2005pg, Gomis:2004pw}.
Likewise, a Dirac equation for spacetime fermions has been obtained from the quantization of a supersymmetric free NR string action~\cite{Kim:2007pc}, and the NR limit of the target space supergravity action has also recently been studied~\cite{Bergshoeff:2021tfn}.
It is of imminent importance to extend these works to map out a more complete picture of NR superstring theories.
This would not only improve our understanding of the DLCQ of string/M-theory, in a formalism that is covariant with respect to NR target-space gauge symmetries, but also enable us to develop a top-down view of NR holographic dualities. {Moreover, it would be intriguing to explore NR superstring amplitudes and generalizing the higher-genus results in \cite{Gomis:2000bd} to superstrings.}

As a concrete example of such NR holographic dualities, the Spin Matrix theory (SMT) limits of AdS/CFT that we discussed in \S\ref{sec:nhd} are expected to provide a fertile testing ground.
Immediate goals include quantizing the SMT string sigma model, including its Galilean Conformal algebra (GCA) of residual reparametrization symmetries (see also~\cite{Kluson:2018egd,Kluson:2021sym}), and obtaining its beta functions.
Since we can deduce the effects of the SMT limit on the AdS geometry, a set of proposed consistent backgrounds already exists.
Also for these SMT strings, the supersymmetrization of the theory is still underdeveloped.
It should be noted that another sigma model with nonrelativistic worldsheet geometry and GCA symmetries exist, which is likewise obtained from a tensionless limit~\cite{Schild:1976vq,Isberg:1993av,Bagchi:2013bga}.
While the resulting strings are fundamentally different, the developments in the quantization~\cite{Bagchi:2020fpr,Bagchi:2021rfw} and supersymmetrization~\cite{Bagchi:2016yyf} of this latter tensionless string can perhaps be of use for SMT strings.
Additionally, it would be very interesting to revisit the NR string theory obtained in \cite{Gomis:2005pg,Sakaguchi:2007ba} from a limit on AdS${}_5 \times S^5$. See for example~\cite{Fontanella:2021hcb,Kluson:2021qqv} for recent discussions of uniform light-cone gauge fixing.

More generally, it would be intriguing to complete the program of building up a duality web in NR string theory, which would provide a new window on studying various nonperturbative sectors in M-theory.
This analysis can be approached both using D$p$-branes as probes (see for example \cite{Ebert:2021mfu}), whose analogs in NR string theory have been discussed in this review, or using supergravity.
For this purpose, it is essential to include Ramond-Ramond charges in the framework of NR string theory and to further explore the different $p$-brane limits generalizing the NR string limit that were discussed in \S\ref{sec:pbl}.
It would also be interesting to understand the relation to the recent construction of NR theories from null reduction of M5-branes~\cite{Lambert:2020zdc,Mouland:2021urv, Kluson:2021djs}.

We have also seen that the T-duality-invariant framework of double field theory (DFT) appears to be particularly suited for studying NR string theory, as it incorporates both the limit and null reduction/duality approach.
As we mentioned in \S\ref{sssec:dft}, the formalism of DFT naturally incorporates several notions of non-Riemannian geometry.
As a result, many DFT-covariant constructions that were originally developed for relativistic string theory can be efficiently applied to NR string theory, including the target-space actions~\cite{Cho:2019ofr,Gallegos:2020egk} and supersymmetric worldsheet actions~\cite{Park:2016sbw,Blair:2019qwi}.
Similar results can be obtained in exceptional field theory~\cite{Berman:2019izh,Blair:2021waq}, and both frameworks are expected to be useful in building a broader understanding of NR strings and their related theories.

So far, we have focused on string theories obtained from NR limits.
However, building on recent work on the NR expansion of general relativity~\cite{VandenBleeken:2017rij,Hansen:2019pkl,Hansen:2020pqs,Ergen:2020yop}, a similar NR expansion of string theory using a small but nonzero Regge slope has been considered~\cite{Hartong:2021ekg}.
This expansion would allow us to consider relativistic corrections to the NR limit  order by order.

There also exist other notions of NR strings and membranes that are not covered by this review. For example, a particle limit of relativistic strings has been studied in \cite{Batlle:2016iel,Gomis:2016zur,Batlle:2017cfa}. This limit gives rise to non-vibrating Galilean strings described by a NR worldsheet, propagating in a target-space geometry that is Newton-like, equipped with a codimension-one foliation structure.
For another example, sigma models with various Lifshitz scalings have been applied to construct the worldvolume theories for membranes \cite{Horava:2008ih} and strings~\cite{Yan:2021xnp}, in the absence of any worldvolume (Lorentzian nor Galilean) boost symmetries.
Such endeavours beyond the relativistic framework may eventually lead to an alternative route towards the unification of different string theories, as well as a larger class of holographic dualities.
Additionally, there exist several `bottom-up' constructions of NR holography, many of which arise from explicit symmetry breaking in relativistic parent theories, see for example the review~\cite{Taylor:2015glc}.

\subsection*{Acknowledgements}

We thank Eric Bergshoeff, Chris Blair, Stephen Ebert, Domingo Gallegos, Umut G\"{u}rsoy, Troels Harmark, Niels Obers, Jeong-Hyuck Park, Peter Schupp and Natale Zinnato for useful discussions and comments on this review.
The work of GO is supported in part by the project ``Towards a deeper understanding of black holes with non-relativistic holography'' of the Independent Research Fund Denmark (grant number DFF-6108-00340) and also by the Villum Foundation Experiment project 00023086.

\bibliographystyle{JHEP}
\bibliography{anrs}

\end{document}